\documentclass{amsart}
\usepackage{newtxtext,newtxmath}       
\usepackage{helvet}         
\usepackage{courier}        
\usepackage{graphicx}        
\usepackage[bottom]{footmisc} 
\usepackage{amsmath,amssymb,amsfonts}        
\usepackage{verbatim, booktabs}
\usepackage{bbm, bm}
\usepackage{algpseudocode}
\usepackage{wrapfig}
\usepackage{caption}
\usepackage{chngcntr}
\usepackage{apptools}
\usepackage[latin1]{inputenc}

\makeindex             

\usepackage{subcaption}
\usepackage{lipsum, enumitem}
\usepackage{epstopdf}
\usepackage{varwidth}
\usepackage{xcolor, color, booktabs}
\usepackage{bbm, bm}

\newcommand{\N}{\mathbb{N}}

\newcommand{\R}{\mathbb{R}}
\newcommand{\bbE}{\mathbb{E}}
\newcommand{\bbV}{\mathbb{V}}
\newcommand{\Expt}{\mathbb{E}}
\newcommand{\cT}{\mathcal{T}}
\newcommand{\cL}{\mathcal{L}}
\newcommand{\bmX}{\bm{X}}
\newcommand{\bmY}{\bm{Y}}
\newcommand{\D}{\mathrm{d}}

\DeclareMathOperator*{\argmin}{argmin}

\theoremstyle{definition}

\newtheorem*{refproof*}{Proof}

\ifpdf
\DeclareGraphicsExtensions{.eps,.pdf,.png,.jpg}
\else
\DeclareGraphicsExtensions{.eps}
\fi

\begin{document}

    \author{Adam Spannaus}
    \address{Department of Mathematics, University of Tennessee,
        Knoxville TN, USA 37996}
    \email{aspannaus@gmail.com}

    \author{Vasileios Maroulas}
    \address{Department of Mathematics, University of Tennessee,
        Knoxville TN, USA 37996}
    \email{maroulas@math.utk.edu}
    
    \author{David J. Keffer}
    \address{Department of Materials Science and Engineering,
        University of Tennessee, Knoxville TN, USA 37996}
    \email{dkeffer@utk.edu}

    \author{Kody J. H. Law}
    \address{School of Mathematics, University of Manchester, UK }
    \email[Corresponding Author]{kodylaw@gmail.com}
    

\title{Bayesian Point Set Registration}
\begin{abstract}
    Point set registration involves identifying a smooth invertible 
transformation between corresponding points in two point sets, 
one of which may be smaller than the other and possibly corrupted by observation noise.
This 
problem is traditionally decomposed into two separate optimization problems: 
(i) assignment or correspondence, and 
(ii) identification of the optimal transformation between the ordered point sets.
In this work, we propose an approach solving both problems simultaneously.
    In particular, a coherent Bayesian formulation of the problem 
    results in a marginal posterior distribution on the transformation, 
    which 
    is explored within
    a Markov chain Monte Carlo scheme. 
    Motivated by Atomic Probe Tomography (APT), 
    in the context of structure inference for high entropy alloys (HEA),
    we focus on the registration of noisy sparse observations 
    of rigid transformations of a known reference configuration.
    Lastly, we test our method on synthetic data sets.
    \end{abstract}

\maketitle


\section{Introduction}\label{sec:1}
 
In recent years, a new class of materials has emerged, called
High Entropy Alloys (HEAs).
The resulting HEAs possess unique mechanical properties and have shown marked
resistance to high-temperature, corrosion, fracture and 
fatigue \cite{gao2016high, zhang2014microstructures}. 
HEAs demonstrate a `cocktail' effect \cite{jien2006recent}, in which the mixing 
of many components results in properties not possessed by any single component individually. 
Although these metals hold great promise for a wide variety of 
applications,
the greatest impediment in tailoring the design of HEAs to specific applications 
is the inability to accurately predict their atomic structure and chemical ordering.
This prevents Materials Science researchers from constructing structure-property
 relationships necessary for targeted materials discovery. 

An important experimental characterization technique used to determine local 
structure of materials at the atomic level is Atomic Probe Tomography (APT) 
\cite{Larson:2013:aptbook,Miller:2013:aptbook}.  APT provides an 
identification of the atom type and its position in space
within the sample.  APT has been successfully applied to the characterization 
of the HEA, AlCoCrCuFeNi \cite{santodonato2015deviation}.
Typically, APT data sets consist of $10^6$ to $10^7$ atoms. Sophisticated reconstruction 
techniques are employed to generate the coordinates based upon the construction 
of the experimental apparatus.  APT data has two main drawbacks: (i) up to 66\% 
of the data is missing
and (ii) the recovered data is corrupted by noise.
The challenge is to uncover the true atomic level structure and chemical ordering 
amid the noise and  missing data, thus giving material scientists an 
unambiguous description of the atomic structure of these novel alloys. 
Ultimately, our goal is to infer the correct spatial alignment
and chemical ordering of a dataset, 
 herein referred to as a configuration, containing up to $10^7$ atoms.
This configuration will be probed by individual registrations of the observed point sets 
in a neighborhood around each atom.

In this paper we outline our approach to this unique registration problem of 
finding the correct chemical ordering
and atomic structure in a noisy and sparse dataset. While we do not
solve the problem in full generality here, we 
present a Bayesian formulation of the model and a general algorithmic approach, 
which allows us to 
confront the problem with a known reference, and can be readily generalized to 
the full problem of an unknown reference.

In Section~\ref{S:2} we describe the problem and our Bayesian
formulation of the statistical model. In Section~\ref{S:3}, we 
describe Hamiltonian Monte Carlo, a sophisticated Markov chain Monte Carlo
technique used to sample from multimodal densities, which we use in
our numerical experiments in Section~\ref{S:4}. Lastly, we conclude
with a summary of the work presented here and directions for 
future research. 

\section{Problem Statement and Statistical Model}\label{S:2}

An alloy consists of a large configuration of atoms, henceforth ``points'', which
are rotated and translated instances of 
a reference collection of points, denoted $\bmX = (X_1, \dots, X_N), X_i\in\R^d$
for $1\leq i\leq N$ which is the matrix representation of the reference points.
The tomographic observation of this configuration is missing some percentage of the points
and is subject to
noise, which is assumed additive and Gaussian.
The sample consists of a single point
and its $M$ nearest neighbors, where $M$ is of the order 10.
If $p \in [0,1]$ is the percent observed, i.e. $p=1$ means all points
are observed and $p=0$ means no points
are observed, then the reference point set will be comprised of $N=\lceil M/p \rceil$ points.
We write the matrix representation of the noisy data point as 
$\bmY = (Y_1, \dots, Y_M), Y_i\in\R^d$, for $1\leq i\leq M$.

The observed points have labels, but the reference points do not.  
We seek to register these noisy and sparse point
sets, onto the reference point set.
The ultimate goal is to identify the ordering 
of the labels of the points (types of atoms) in a configuration. 
We will find the best assignment and rigid transformation
between the observed
point set and the reference point set.
Having completed the registration process for all observations
in the configuration, we
may then construct a three dimensional distribution of labeled points 
around each reference point,
and the distribution of atomic composition is readily obtained.  

The point-set registration problem has two crucial elements.
The first is the correspondence, or assignment of each point in the observed 
set to the reference set.  
The second is the identification of the optimal transformation from within an appropriate class of transformations.
If the transformation class is taken to be the rigid transformations, 
then each of the individual problems is easily solved by itself, 
and naive methods simply alternate the solution of each individually until convergence. 

One of the most frequently used point set registration algorithms is the iterative closest point method, which alternates
between identifying the optimal transformation for a given correspondence, and then corresponding 
closest points \cite{BeslMcKay}.  If the transformation is rigid, then both problems are uniquely solvable.
If instead we replace the naive closest point strategy 
with the assignment problem, so that any two observed points correspond to two different reference points, 
then again the problem can solved with a linear program \cite{li20073d}.
However, when these two solvable problems are combined 
into one, the resulting problem is non-convex \cite{papazov2011stochastic},
and no longer admits a unique solution, even for the case of rigid
transformations as considered here.
The same strategy has been proposed with 
more general non-rigid transformations \cite{chui2003new}, 
where identification of the optimal transformation is no longer analytically solvable.
The method in \cite{myronenko2010point} 
minimizes an upper bound on their objective function, and 
is thus also susceptible to getting stuck in a local
basin of attraction. 
We instead take a Bayesian formulation of the problem that
will simultaneously find the transformation
and correspondence between point sets. Most
importantly, it is designed to avoid local basins of attraction and locate a global minimum.

\begin{wrapfigure}{r}{0.5\textwidth}
    \begin{center}
        \includegraphics[width=0.48\textwidth]{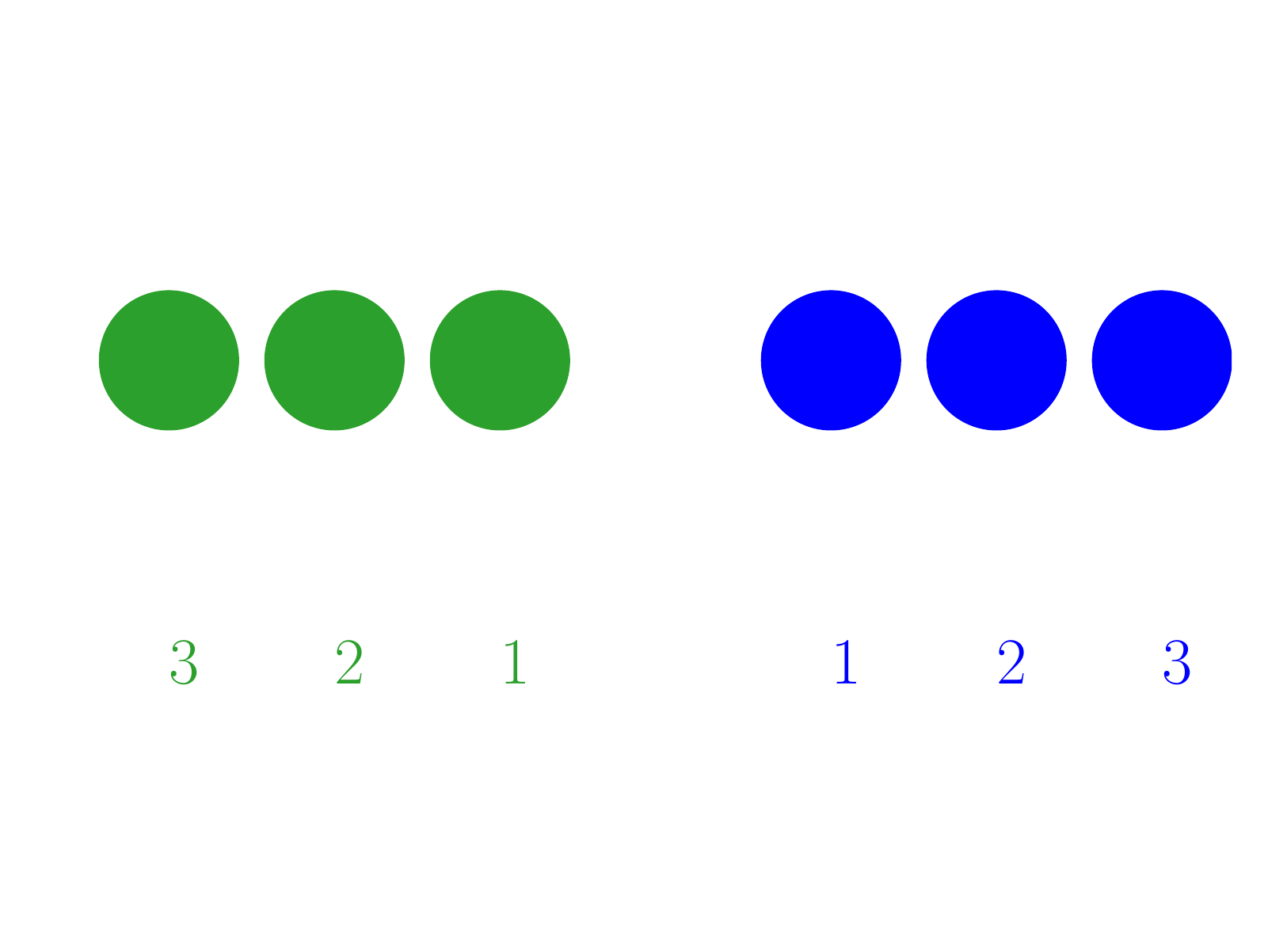}
    \end{center}
    \caption{Setup for incorrect registration; alternating assignment and $\ell^2$ minimization}\label{fig:reg}
\end{wrapfigure}

We will show how alternating
between finding correspondences and minimizing distances can lead to
an incorrect registration. 
Consider now the setup in Fig.~(\ref{fig:reg}). 
If we correspond closest points first, then all three green points would
be assigned to the blue `1'. 
Then, identifying the single rigid transformation to minimize the distances  
 between all
three green and the blue `1' would 
yield a local minimum, with no correct assignments. 
If we consider instead {\em assignments}, so that no two observation points can correspond to the same reference point,
then again it is easy then to see two equivalent solutions with the eye.  The first 
is a pure translation, and the second can be obtained for example by
one of two equivalent rotations around the mid-point between `1's, by $\pi$ or $-\pi$.
The first only gets the assignment of `2' correct, while the second is correct.  
Note that in reality the reference labels are unknown, so both are equivalent for us. 

Here it is clear what the solutions are, 
but once the problem grows in scale, the answer is not always so clear.
This simple illustration of degenerate (equal energy) multi-modality of the registration objective function
arises from physical symmetry of the reference point-set.  This will be an important consideration 
for our reference point sets, which will arise as a unit cell of a lattice, 
hence with appropriate symmetry.  
We will never be able to know the registration beyond these symmetries, 
but this will nonetheless not be the cause of concern, as symmetric solutions will be considered equivalent.
The troublesome multi-modality arises in the presence of noisy and partially observed point sets,
where there may be local minima with higher energy than the global minima.

The multi-modality of the combined problem, in addition to the limited information in
the noisy and sparse observations, motivates the need for a global 
probabilistic notion of solution for this problem.  It
is illustrated in the following subsection that the problem 
lends itself naturally to a flexible Bayesian formulation which circumvents the intrinsic
shortcomings of deterministic optimization approaches for non-convex problems.
Indeed at an additional computational cost, we
obtain a distribution of solutions, rather than a point estimate,
so that general quantities of interest are estimated and uncertainty
is quantified. 
In case a single point estimate is required we define
an appropriate optimal one (for example the global energy minimizer or probability maximizer).

\subsection{Bayesian Formulation}\label{S:21}

We seek to compute the registration between the 
observation set and reference set.  
We are concerned primarily with rigid transformations of the form 
\begin{equation}
    \mathcal{T}(X; \theta) = R_\theta X + t_\theta,\label{eq:rigid_trans}
\end{equation}
where $R_\theta \in \R^{d\times d}$ is a rotation 
and $t_\theta \in \R^d$ is a translation vector.

Write  $[\mathbb{T}(\bmX; \theta)]_{ki} = \cT_k(X_i)$ for $1\leq i \leq N$, 
$1 \leq k \leq d$, and where $X_i$ is the i$^{th}$ column of $\bmX$. Now let $\xi\in \R^{d\times M}$ with entries 
$\xi_{ij} \sim N(0,\gamma^2)$, and assume the following statistical model
\begin{equation}
\bmY = \mathbb{T}(\bmX;\theta)C + \xi, \label{eqn:mapping} 
\end{equation}
for $\xi, \theta$ and $C$ independent.

The matrix of correspondences $C\in\{0,1\}^{N\times M}$, is such that
$\sum_{i=1}^N\,C_{ij} = 1, 1\leq j\leq M$, and each observation point 
corresponds to only one reference point. So if $X_i$ matches $Y_j$ then $C_{ij} = 1$, otherwise, $C_{ij} = 0$. We let $C$ 
be endowed with a prior, $\pi_0(C_{ij} = 1) = \pi_{ij}$ for $1\leq i\leq N$ and $1\leq j\leq M$.
Furthermore, assume a prior on the transformation parameter $\theta$ given 
by $\pi_0(\theta)$.
The posterior distribution then takes the form
\begin{equation}\label{eq:joint_post}
\pi(C,\theta\mid\bmX,\bmY)\propto \cL(\bmY\mid\bmX, C, \theta)\pi_0(C)\pi_0(\theta),
\end{equation}
where $\cL$ is the likelihood function associated with Eqn.~\eqref{eq:rigid_trans}.

For a given $\tilde{\theta}$, 
an estimate $\hat C$ can be constructed \textit{a posteriori}  
by letting $\hat C_{i^*(j),j} = 1$ for $j=1,\dots, M$ and zero otherwise, where 
\begin{equation}
i^*(j) = \argmin_{1\leq i\leq N} |Y_j - \cT(X_{i}; \tilde{\theta})|^2 \,.
\end{equation}
For example, $\tilde{\theta}$ may be taken as the maximum a posteriori (MAP) estimator or the mean.
We note that $\hat{C}$ can be constructed either with a closest point approach, or
via assignment to avoid multiple registered points assigned to the same reference.

Lastly, we assume the $j^{th}$ observation only depends on the $j^{th}$ 
column of the correspondence matrix, and so
$Y_i, Y_j$ are conditionally independent with respect to the matrix $C$
for $i\neq j$.
This does not exclude the case where
multiple observation points 
are assigned to the same reference point, 
but as mentioned above such scenario should have zero probability.

To that end, instead of considering the full joint posterior in 
Eqn.~\eqref{eq:joint_post} we will
focus on the marginal of the transformation
\begin{equation}\label{eq:marg_post}
\pi(\theta\mid\bmX,\bmY)\propto \cL(\bmY\mid\bmX, \theta)\pi_0(\theta).
\end{equation} 

Let $C_{j}$ denote the $j^{th}$ column of $C$.
Since $C_{j}$ is completely determined by the single index $i$ at which it takes the value 1,
the marginal likelihood takes the form 
\begin{align}\nonumber
    \sum_{C} p(Y_j\mid \bmX,\theta, C )\pi_0(C) &= \sum_{i=1}^N p(Y_j\mid \bmX,\theta,C_{ij}=1)\pi_0(C_{ij}=1)\\
	\nonumber
        &= \sum_{i=1}^N \pi_{ij}p(Y_j\mid \bmX,\theta,C_{ij}=1)\\
        &\propto \pi_{ij}\exp\left\{-\frac1{2\gamma^2} |Y_j - T(X_i;\theta)|^2 \right\} \, .
\end{align}

The above marginal together with the conditional independence assumption 
allows us to construct the likelihood
function of the marginal
posterior, Eqn.~\eqref{eq:marg_post}, as follows 
\begin{align}\nonumber\label{eqn:liklihood}
\cL(\bmY\mid \bmX, \theta) &= \prod_{j=1}^M\,p(Y_j\mid \bmX, \theta) \\
    &\propto \prod_{j=1}^M\sum_{i=1}^N\,\pi_{ij}\exp\left\{-\frac1{2\gamma^2} |Y_j - T(X_i;\theta)|^2 \right\}\, .
\end{align}

Thus the posterior in question is
\begin{align}\nonumber\label{eqn:posterior}
\pi(\theta\mid\bmX, \bmY)&\propto \cL(\bmY\mid\bmX, \theta)\pi_0(\theta)\\
    &=\prod_{j=1}^M\sum_{i=1}^N\,\pi_{ij}\exp\left\{-\frac1{2\gamma^2} |Y_j - T(X_i;\theta)|^2 \right\}\pi_0(\theta)\, .
\end{align}

At its heart, point set registration is an optimization problem. 
Consider a prior on $\theta$
such that $\pi_0(\theta)\propto \exp(-\lambda R(\theta))$, 
where $\lambda>0$. Then we have the following 
objective function
\begin{equation}
E(\theta) = -\sum_{j=1}^M\log\sum_{i=1}^N\,\pi_{ij}\exp\left\{-\frac1{2\gamma^2} |Y_j - T(X_i;\theta)|^2 \right\} + \lambda R(\theta)\, .\label{eqn:obj_fun}
\end{equation}

The minimizer, $\theta^*$, of the above, Eqn.~\eqref{eqn:obj_fun} is also the maximizer of a
posteriori probability under Eqn.~\eqref{eqn:posterior}.   
It is called the maximum a posteriori estimator.
This can also be viewed as maximum likelihood estimation regularized by $\lambda R(\theta)$.
 
By sampling consistently from the posterior, we may estimate
quantities of interest, such as 
moments, together with quantified uncertainty.
Additionally, we may recover other point estimators, 
such as local and global modes.

\section{Hamiltonian Monte Carlo}\label{S:3}
 
Monte Carlo Markov chain (MCMC) methods are a natural choice for sampling from 
distributions which can be evaluated pointwise up to a normalizing constant, such as 
the posterior Eqn.~\eqref{eqn:posterior}.  
Furthermore, MCMC comprises the workhorse of Bayesian computation, 
often appearing as crucial components of more sophisticated sampling algorithms.
Formally, an MCMC 
simulates a distribution $\mu$ over a state space $\Omega$ 
by producing an ergodic Markov chain $\{w_k\}_{k\in\N}$ that has $\mu$ as
its invariant distribution, i.e. 
\begin{equation}
    \frac1{K}\sum_{k=1}^K\,g(w_k) \rightarrow \int_{\Omega}\,g(w)\mu(dw) = \Expt_{\mu}g(w)\, ,
\end{equation}
with probability 1, for $g\in L^1(\Omega)$.

The Metropolis-Hastings method is a general MCMC method defined
by choosing $\theta_0 \in \rm{supp}(\pi)$ 
and iterating the following two steps for $k\geq 0$
\begin{itemize}
    \item[(1)] Propose: $\theta^* \sim Q(\theta_{k},\cdot)$.
    \item[(2)] Accept/reject: Let $\theta_{k+1}=\theta^*$ with probability 
    \[
    \alpha(\theta_k,\theta^*)=\min\left\{1,\frac{\pi(\theta^*)Q(\theta^*,\theta_{k})}
        {\pi(\theta_k)Q(\theta_{k},\theta^*)}\right\} \, ,
    \]
    and $\theta_{k+1}=\theta_k$ otherwise.
\end{itemize}

In general, random-walk proposals $Q$ can 
result in MCMC chains which are
slow to explore the state space and susceptible to
getting stuck in local basins of attraction. 
Hamiltonian Monte Carlo (HMC) is designed to improve this shortcoming. 
HMC is a Metropolis-Hastings method \cite{duane1987hybrid, Neal1996} which 
incorporates gradient information of the log density with a simulation
of Hamiltonian dynamics to efficiently explore the state space 
and accept large moves of the Markov chain.  
Heuristically, the gradient yields $d$ pieces of information, for a $\R^d$-valued
variable and scalar objective function, as compared with one piece of information from
the objective function alone. Our description here of the HMC algorithm follows
that of \cite{brooks2011handbook} and the necessary foundations of Hamiltonian 
dynamics for the method can be found in \cite{teschl2012ordinary}.

Our objective here is to sample from a specific target density
\begin{equation}\label{eq:target}
\pi(\theta) \propto \exp\{- E(\theta)\}\, 
\end{equation}
over $\theta$, where $E(\theta)$ is as defined in Eqn.~\eqref{eqn:obj_fun} and $\pi(\theta)$ is
of the form given by Eqn.~\eqref{eqn:posterior}.

First, an artificial momentum variable $p \sim N(0, \Gamma)$, independent of $\theta$,
is included into Eqn.~\eqref{eq:target}, for a symmetric positive definite mass matrix $\Gamma$, that is usually a scalar multiple
of the identity matrix.
Define a Hamiltonian now by 
\[
H(p, \theta) = E(\theta) + \frac1{2}p^T\Gamma^{-1} p
\]
where $E(\theta)$ is the ``potential energy'' and $\frac1{2}p^T\Gamma^{-1}p$
is the ``kinetic energy''.
 
Hamilton's equations of motion for $p,\theta\in \R^d$ are, for $i=1,\dots, d$ :
\begin{align*}
\frac{\D\theta_i}{\D t} &= \frac{\partial{H}}{\partial{p_i}}\\
\frac{\D p_i}{\D t} &= -\frac{\partial{H}}{\partial{\theta_i}}
\end{align*}

In practice, the algorithm creates a Markov chain on the joint position-momentum space $\R^{2d}$,
by alternating between independently sampling from the marginal Gaussian on momentum $p$, 
and numerical integration of Hamiltonian dynamics along an energy contour 
to update the position. 
If the initial condition $\theta \sim \pi$
and we were able to perfectly simulate the dynamics, 
this would give samples from $\pi$
because the Hamiltonian $H$ remains constant along trajectories. 
Due to errors in numerical approximation, the value of $H$ will vary.
To ensure the samples are indeed drawn from the correct distribution,
a Metropolis-Hastings accept/reject step is incorporated into the method.

In particular, after a new momentum is sampled, suppose the chain is in the state $(p,\theta)$.
Provided the numerical integrator is reversible,
the probability of accepting the proposed point $(p^*, \theta^*)$ 
takes the form
\begin{equation}\label{eq:hmcacc}
    \alpha((p, \theta),(p^*, \theta^*)) = \min\left\{1, \exp\left\{ H(p, \theta) - H(p^*, \theta^*) \right\} \right\}.
\end{equation}
If $(p^*, \theta^*)$ is rejected, the next state remains unchanged from the previous
iteration.  However, note that a fresh momentum variable is drawn each step, 
so only $\theta$ remains fixed.  
Indeed the momentum variables can be discarded, as they are only auxiliary variables.
To be concrete, the algorithm requires an initial state $\theta_{0}$, 
a reversible numerical integrator, integration step-size $h$, and number of steps $L$.  
Note that reversibility of the integrator is crucial such that the proposal integration 
$Q((p, \theta),(p^*, \theta^*))$ is symmetric and drops out of the acceptance probability
in Eqn.~\eqref{eq:hmcacc}.
The parameters $h$ and $L$ are tuning parameters, and are described in detail
\cite{brooks2011handbook, Neal1996}.

The HMC algorithm then proceeds as follows:
\begin{algorithmic}
    \For {$k\geq 0$} \textbf{HMC}:
    \State $p_{k} \gets \xi$ for $\xi\sim \mathcal{N}(0, \Gamma)$
    \Function{Integrator}{$p_k, \theta_k, h$}
    \Return $(p^*, \theta^*)$
    \EndFunction
    \State {$\alpha\gets\min\left\{1, \exp\left\{ H(p_k, \theta_k) - H(p^*, \theta^*) \right\} \right\}$}
    \State $\theta_{k+1}\gets \theta^*$ with probability $\alpha$\textbf{ otherwise}
    \State $\theta_{k+1}\gets \theta_{k}$
    \EndFor
\end{algorithmic}

Under appropriate assumptions \cite{Neal1996}, this method will provide samples $\theta_k \sim \pi$, 
such that for bounded $g: \R^n \rightarrow \R$
\[
\frac1K \sum_{k=1}^K g(\theta_k) \rightarrow
\int_{\R^n} g(\theta)d\theta \quad  \mathrm{as} \quad K\rightarrow \infty \, .
\]

\section{Numerical Experiments}\label{S:4}

To illustrate our approach, we consider numerical experiments on synthetic datasets 
in $\R^2$ and $\R^3$, with varying levels of noise and percentage of observed data.
We focus our attention 
to rigid transformations of the form Eqn.~\eqref{eq:rigid_trans}.

For all examples here, the $M$ observation points are simulated 
as $Y_i \sim N(R_\varphi X_{j(i)} + t,\gamma^2 I_d)$, for a rotation matrix
$R_\varphi$ parameterized by $\varphi$, and some 
$t$ and $\gamma$.  So, $\theta=(\varphi,t)$.  
To simulate the unknown correspondence
between the reference and observation points, for each $i=1,\dots,M$, 
the corresponding index $j(i) \in [1,\dots,N]$ is chosen randomly and without replacement.
Recall that we define percentage of observed points here as 
$p=\frac{M}{N} \in[0,1]$.
We tested various percentages of observed data and 
noise $\gamma$ 
on the observation set, then computed the mean square error (MSE), given 
by Eqn.~\eqref{eqn:error_metric},
between the reference points and the registered observed points,
\begin{equation}
\mathcal{E}(\theta) = \frac1{M}\sum_{i=1}^M \min_{X\in{\bf X}}|R_{\varphi}^T(Y_i - t) - X|^2\, .
\label{eqn:error_metric}
\end{equation}

\begin{figure}
        \begin{minipage}[l]{0.45\textwidth}
            \includegraphics[width=\linewidth]{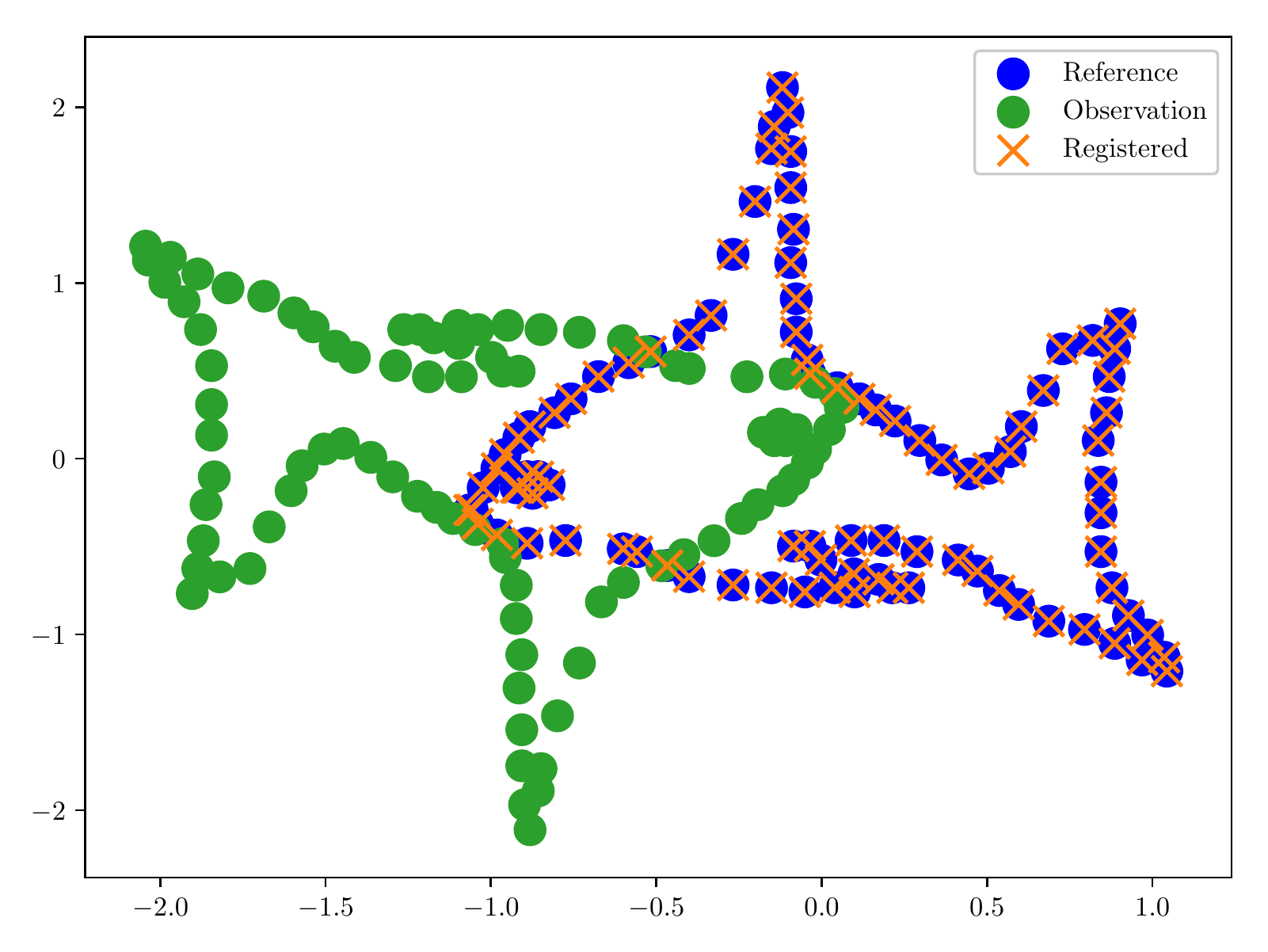}
            \caption{Full data}\label{fig:full_fish}
        \end{minipage}
    \hspace{0.5em}
        \begin{minipage}[r]{0.45\textwidth}
            \includegraphics[width=\linewidth]{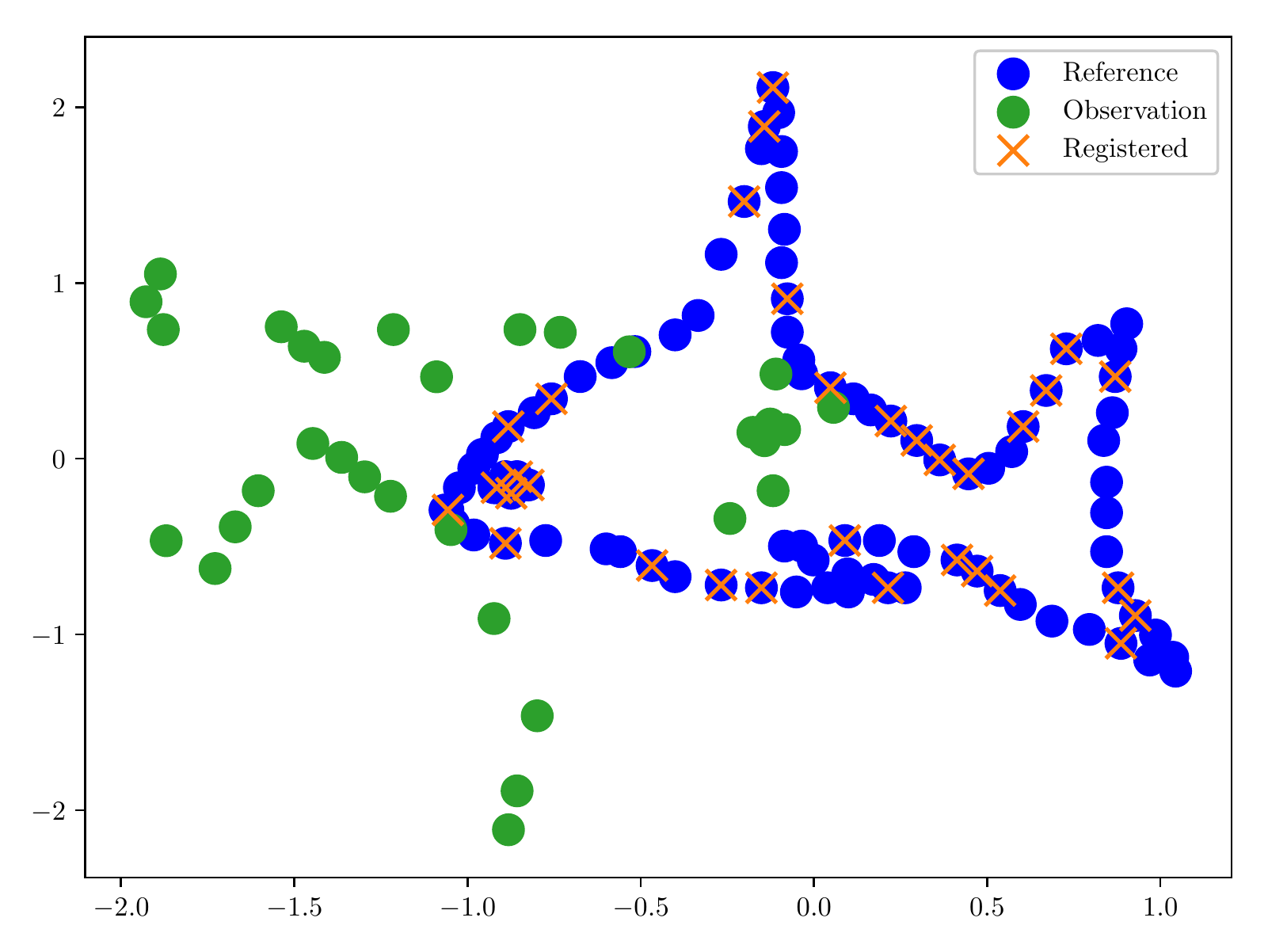}
            \caption{33\% Observed Data}\label{fig:missing_fish}
        \end{minipage}
\end{figure}

\subsection{Two Dimensional Registration}

First we consider noise-free data, i.e. $\gamma=0$ 
(however in the reconstruction some small $\gamma>0$ is used).
The completed registration for the 2-dimensional `fish' set is shown in
Figs.~(\ref{fig:full_fish}, \ref{fig:missing_fish}). The `fish' set is 
a standard benchmark test case for registration algorithms in $\R^2$
\cite{jian2011robust,myronenko2007}. Our
methodology, employing the HMC sampler described in Sect.~\ref{S:3} 
allows for a correct registration, even in the case
where we have only 33\% of the initial data, see Fig.~(\ref{fig:missing_fish}). 

\begin{figure}
    \centering
    \includegraphics[scale=0.4]{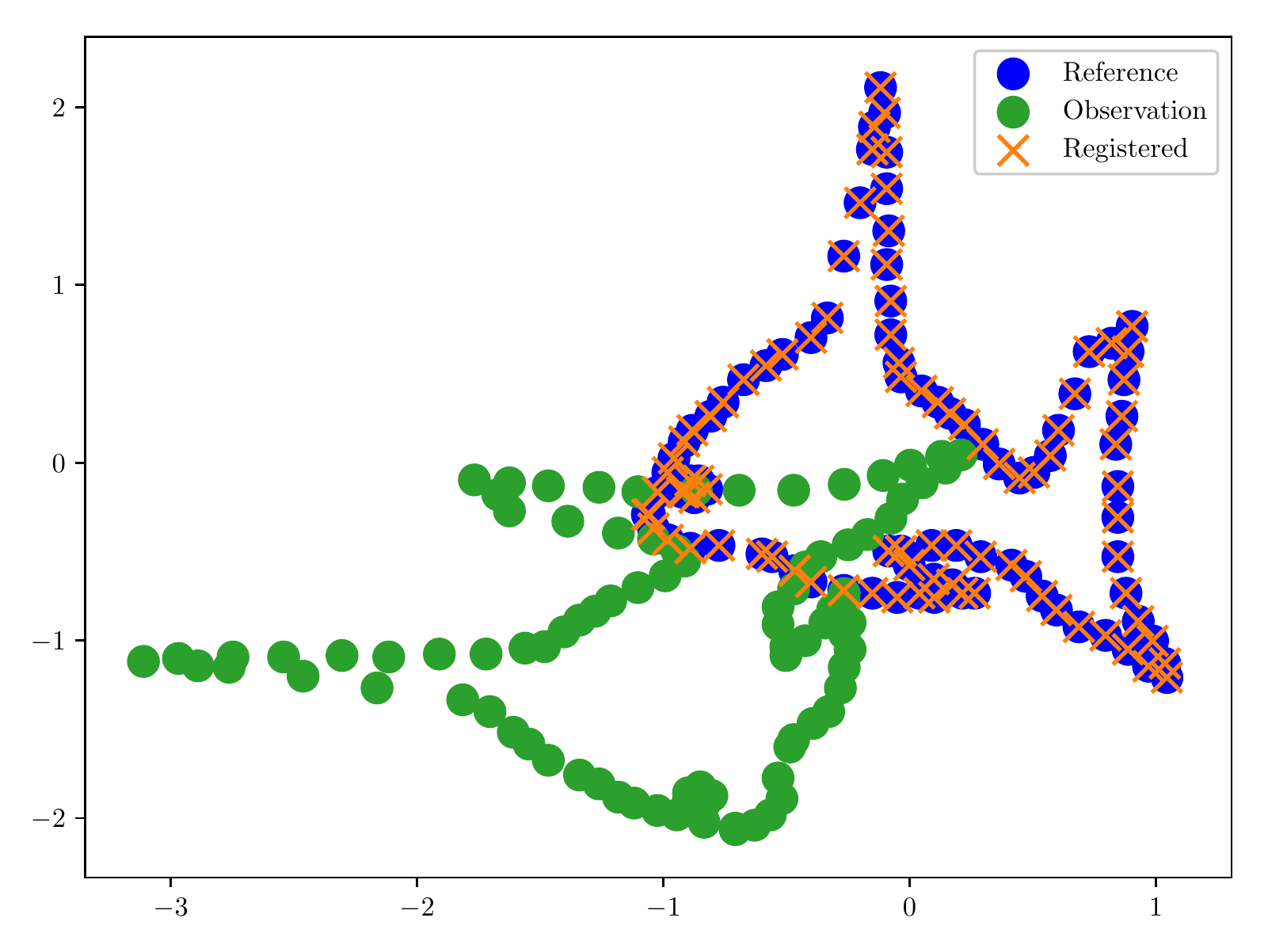}
    \captionof{figure}{Full data, $\gamma=0.5$, average of 25 Registrations.}\label{fig:sample_fish}
\end{figure}

As a final experiment with the `fish' dataset, we took 25 independent identically 
distributed (i.i.d.) realizations of the reference, all having the same transformation, noise,
and percent observed. Since we have formulated the solution of our
registration problem as a density, we may compute moments, and
find other quantities of interest. In this experiment 
we evaluate $\bar{\theta} = \frac1{25}\sum_{k=1}^{25}\hat{\theta}_k$,
where $\hat{\theta}$ is 
our MAP estimator of $\theta$ from the HMC algorithm. 
We then evaluated the transformation under
$\bar{\theta}$. The completed registration is shown in Fig.~(\ref{fig:sample_fish}).
With a relatively small number of configurations, we are able
to accurately reconstruct the data, despite the noisy observations.

\subsection{Synthetic APT Data}

The datasets from APT experiments are perturbed by 
additive noise on each of the points. The variance of this additive noise
is not known in general, and so in practice it should be taken as a hyper-parameter, 
endowed with a hyper-prior, and inferred or optimized.
It is known that the size of the displacement on the order of several \r{A} (Angstroms), 
so that provides a good basis for choice of hyper prior.
In order to simulate this uncertainty in our experiments,
we incorporated additive noise in the form of a truncated Gaussian, 
to keep all the mass within several \r{A} . 
The experiments consider a range of variances in order to measure
the impact of noise on our registration process. 

\begin{figure}
    \centering
    \begin{minipage}{.32\textwidth}
        \includegraphics[width=.32\textwidth]{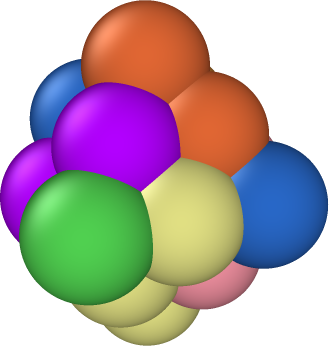}
    \end{minipage}
    \hfill%
    \begin{minipage}{.32\textwidth}
        \includegraphics[width=.32\textwidth]{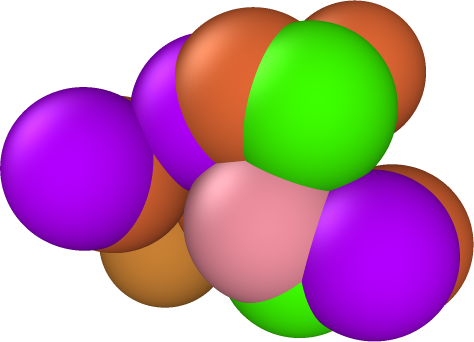}
    \end{minipage}
    \hfill%
    \begin{subfigure}{.32\textwidth}
        \includegraphics[width=.32\textwidth]{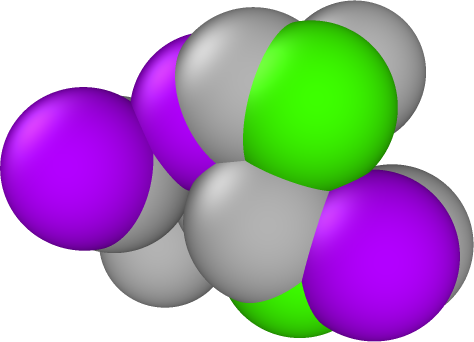}
    \end{subfigure}
        \caption{Example APT data: Left: Hidden truth, Center: Noise added, Right: Missing atoms colored grey.}
        \label{fig:ex_cells}
\end{figure}

In our initial experiments with synthetic data, we have chosen percentages of observed data and
additive noise similar to what Materials Scientist experimentalists have reported in their 
APT datasets. The percent observed of these experimental datasets is approximately
33\%.
The added noise of these APT datasets is harder to quantify. Empirically,
we expect the noise to be Gaussian in form, truncated to be within
1-3 \r{A}. The standard deviation of the added noise is less well-known,
so we will work with different values to asses the method's performance. 
With respect to the size of the cell, a 
displacement of 3\r{A} is significant.
Consider the cell representing the hidden truth in Fig.~(\ref{fig:ex_cells}). The 
distance between the front left and right corners is on the scale of 
3\r{A}.  Consequently a standard deviation of 0.5 for the additive noise
represents a significant displacement of the atoms.

As a visual example, the images in Fig.~(\ref{fig:ex_cells})
are our synthetic test data used to simulate
the noise and missing data from the APT datasets. The leftmost image in 
Fig.~(\ref{fig:ex_cells}) is the hidden truth we seek to uncover. The middle image
is the first with noise added to the atom positions. Lastly, in the right-most
image we have `ghosted' some atoms, by coloring them grey, to give a better visual 
representation of the missing data. In these representations of HEAs, a 
color different from grey denotes a distinct type of atom. What we seek is to infer the 
chemical ordering and atomic structure of the left image, from transformed versions of
the right, where $\gamma= 0.5$.

For our initial numerical experiments with simulated APT data, we choose
a single reference and observation, and consider two different percentages 
of observed data, 75\% and 45\%. For both levels 
of observations in the data, we looked at results with three different levels of
added noise on the atomic positions: no noise, and Gaussian noise with
standard deviation of 0.25 and 0.5. The MSE of the processes 
are shown in Table \ref{tab:2}. We initially observe the method is able,
within an appreciably small tolerance, find the exact
parameter $\theta$ in the case of no noise, with both percentages of observed data. 
In the other cases, as expected, the error scales with the noise. This follows
from our model, as we are considering a rigid transformation between the observation
and reference, which is a volume preserving transformation.  
If the exact transformation is used with an infinite number of points, 
then the RMSE (square root of Eqn.~\eqref{eqn:error_metric}) is $\gamma$.
 
Now we make the simplifying assumption
that the entire configuration corresponds to the same reference,  
and each observation 
in the configuration corresponds to the same transformation applied to the reference, 
with i.i.d.~noise added to it.
This enables us to approximate the mean and variance of Eqn.~\eqref{eqn:error_metric} 
over these observation realizations, i.e. we obtain a collection $\{\mathcal{E}^l(\theta^l)\}_{l=1}^L$
of errors, where $\mathcal{E}^l(\theta^l)$ is the MSE corresponding to replacing ${\bf Y}^l$ and its
estimated registration parameters $\theta^l$ into Eqn.~\eqref{eqn:error_metric}, where $L$ is 
the total number of completed registrations.
The statistics of this collection of values provide robust estimates of the expected error for a single such registration,
and the variance we can expect over realizations of the observational noise. In other words
\begin{equation}\label{eq:aver}
    \bbE^L \mathcal{E}(\theta) := \frac1L \sum_{l=1}^L \mathcal{E}^l(\theta^l) \, 
\text{ and }\,
    \bbV^L \mathcal{E}(\theta) := \frac1L \sum_{l=1}^L (\mathcal{E}^l(\theta^l) - \bbE^L \mathcal{E}(\theta))^2 \, .
\end{equation}
We have confidence intervals as well, corresponding to a central limit theorem approximation based on these $L$ samples.

\begin{figure}
    \centering
    \begin{minipage}[l]{0.45\textwidth}
            \includegraphics[width=\linewidth]{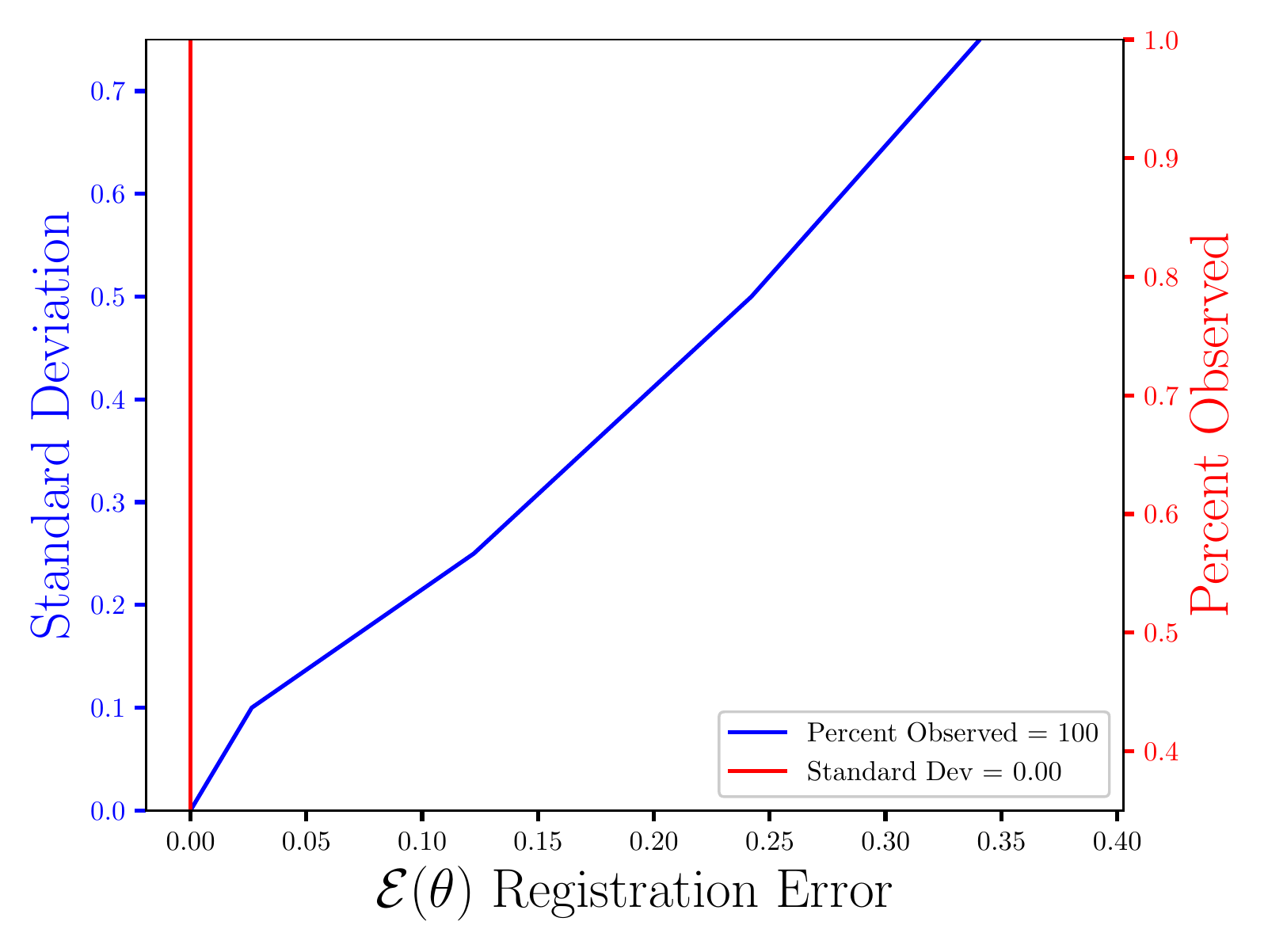}
            \caption{Blue: Full data, Red: Noiseless data}\label{fig:error1}
        \end{minipage}
        \hspace{0.5em}
        \begin{minipage}[r]{0.45\textwidth}
            \includegraphics[width=\linewidth]{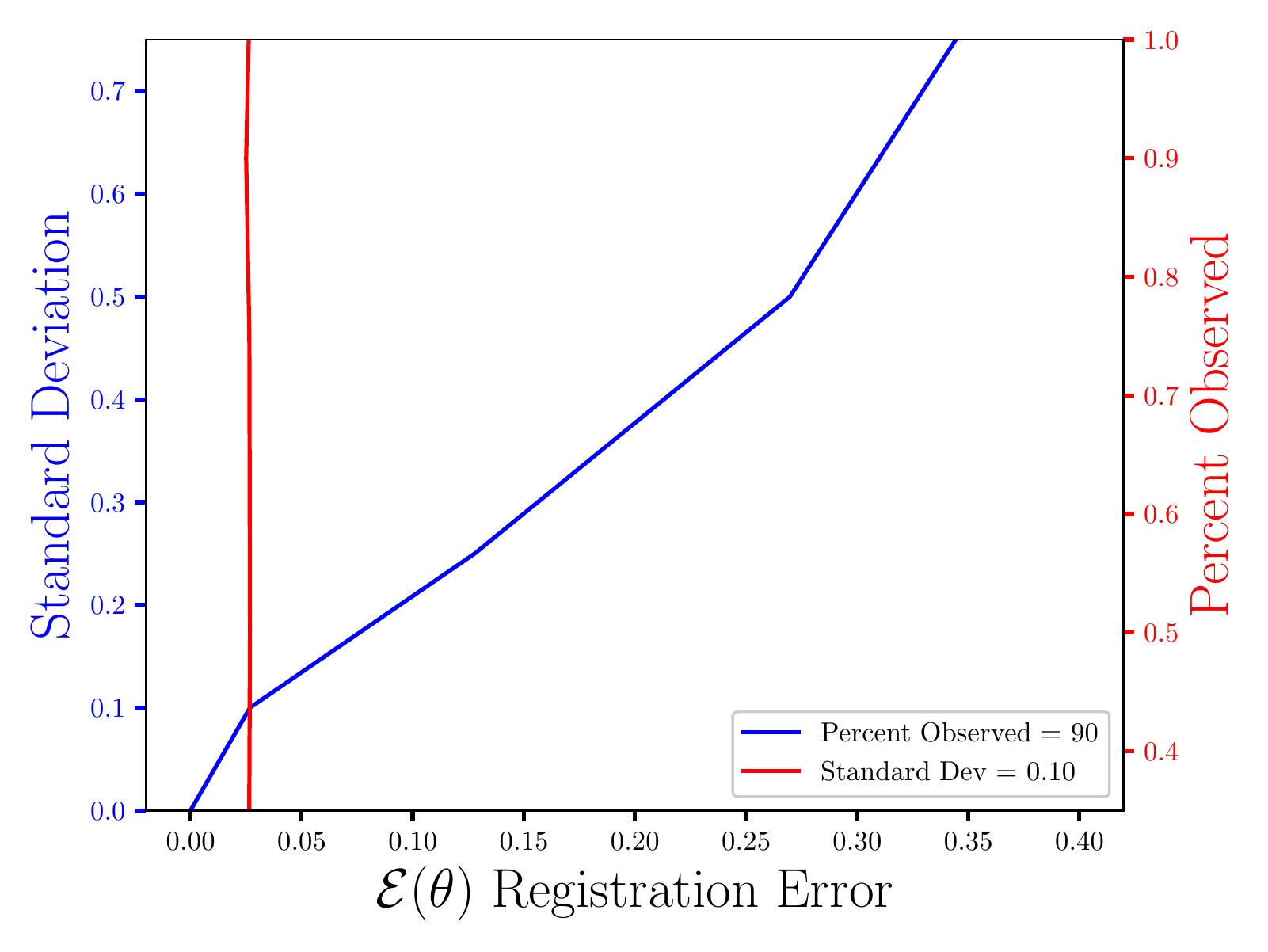}
            \caption{Blue: 90\% Observed, Red: $\gamma = 0.1$}\label{fig:error2}
        \end{minipage}
\end{figure}

\begin{figure}
    \centering
        \begin{minipage}[t]{0.45\textwidth}
            \includegraphics[width=\linewidth]{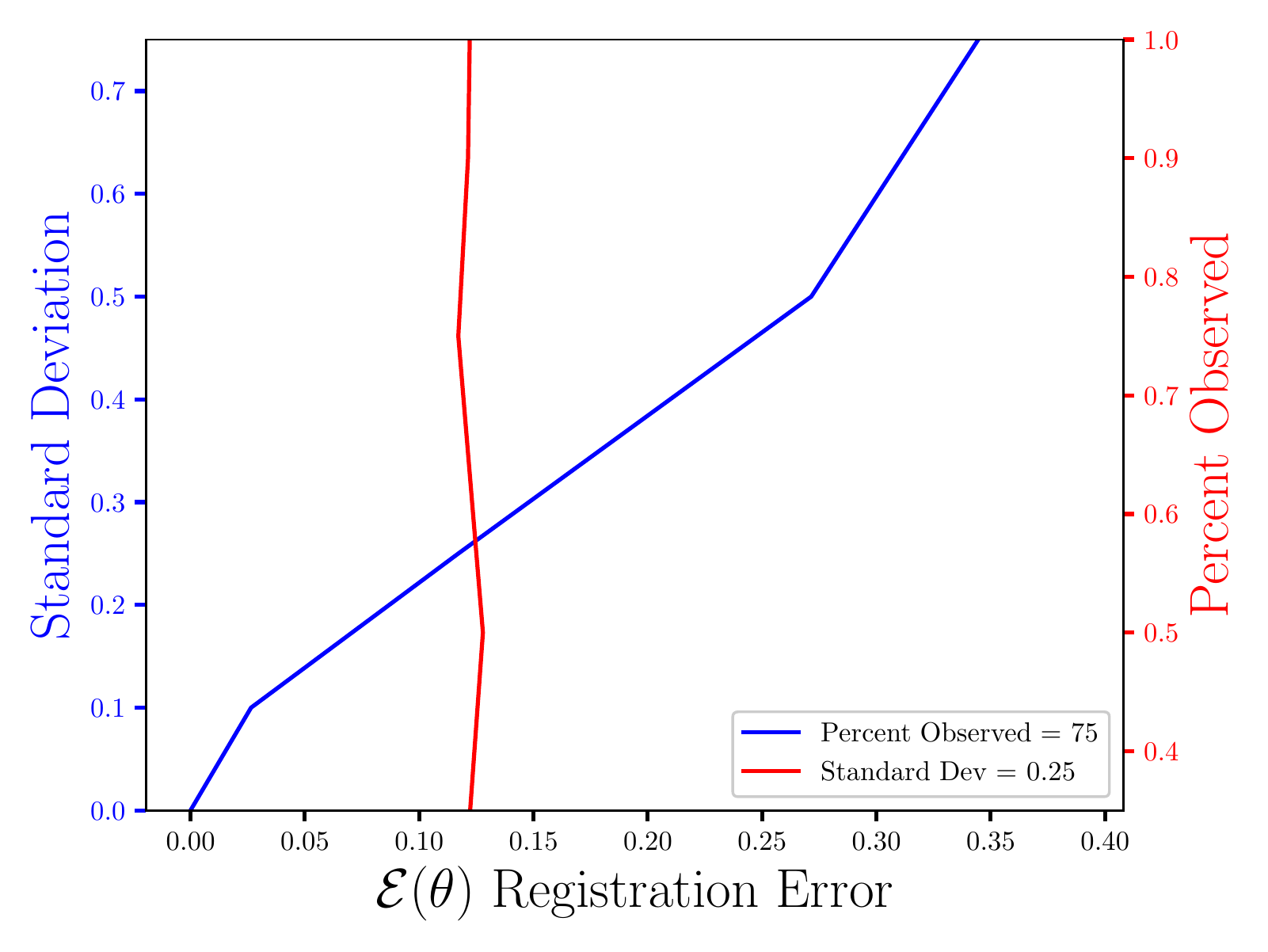}
            \caption{Blue: 75\% Observed, Red: $\gamma = 0.25$}\label{fig:error3}
        \end{minipage}
        \hspace{0.5em}%
        \begin{minipage}[t]{0.45\textwidth}
            \includegraphics[width=\linewidth]{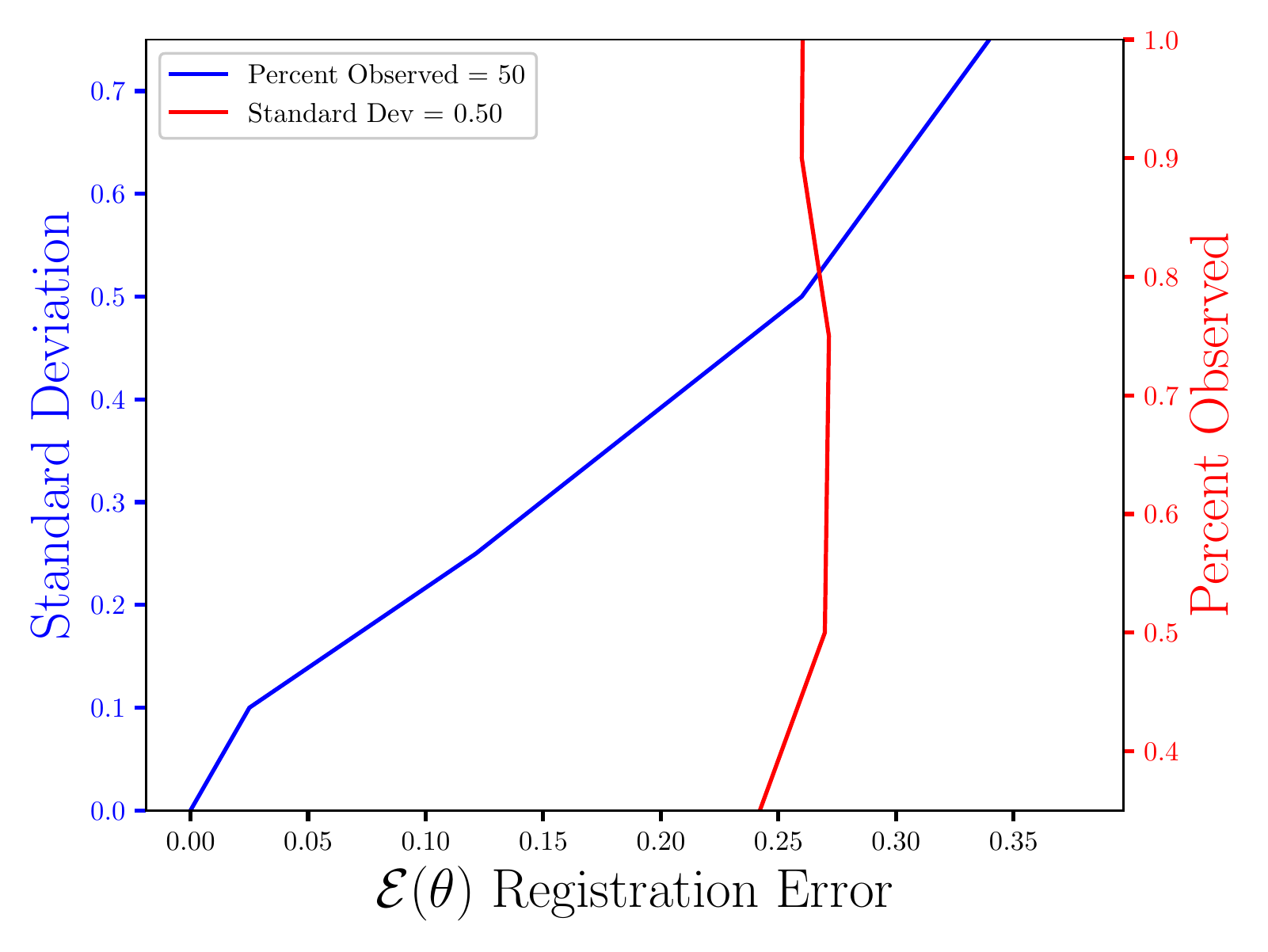}
            \caption{Blue: 50\% Observed, Red: $\gamma = 0.5$}\label{fig:error4}
        \end{minipage}
\end{figure}

In Figs.~(\ref{fig:error1} - \ref{fig:error4}) we computed the registration for $L=125$ i.i.d.~
observation sets corresponding to the same reference, for each combination of 
noise and percent observed data. 
We then averaged all 125 registration errors for a fixed noise/percent observed 
combination, as in Eqn.~\eqref{eq:aver}, and compared the values.
What we observe in Figs.~(\ref{fig:error1} - \ref{fig:error4}) is the registration error scaling with
the noise, which is expected.
What is interesting to note here is that the registration error is essentially constant with
respect to the percentage of observed data, for a fixed standard deviation
of the noise. 
More information will lead to a lower variance in the posterior on the transformation $\theta$,
following from standard statistical intuition.
However, the important point to note is that, as mentioned above, 
for exact transformation, and infinite points, Eqn.~\eqref{eqn:error_metric} will equal $\gamma^2$.
So, for sufficiently accurate transformation, one can expect a sample approximation thereof.
Sufficient accuracy is found here with very few observed points, which is reasonable considering that 
in the zero noise case 2 points is sufficient to fit the 6 parameters exactly.

The MSE registration errors shown in Figs.~(\ref{fig:error1} - \ref{fig:error4}), 
show the error remains essentially constant with respect to the percent observed. Consequently,
if we consider only Fig.~(\ref{fig:error2}), we observe that the blue and red lines
intersect, when the blue has a standard deviation of 0.1, and the associated MSE is 
approximately 0.05. This same error estimate holds for all tested percentages
of observed data having a standard deviation of 0.1. Similar results hold for other combinations 
of noise and percent observed, when the noise is fixed.

\begin{table}
    \centering
    \begin{tabular}{p{1.5cm}p{1.5cm}p{3.2cm}}
        \hline\noalign{\smallskip}
        Standard Deviation & Percent Observed & Registration Error \\
        \noalign{\smallskip}\hline\noalign{\smallskip}
        0.0 & 75\% & $3.49368609883352\mathrm{e}{-11}$\\
        0.0 & 45\% & $4.40071892313178\mathrm{e}{-11}$\\
        0.25 &  75\% & 	0.1702529649951198 \\
        0.25 & 45\% & 	0.1221555853433331\\
        0.5 & 75\% & 0.3445684328735114 \\
        0.5 & 45\% & 0.3643178111314804\\
        \noalign{\smallskip}\hline\noalign{\smallskip}
    \end{tabular}
    \caption{$\mathcal{E}(\theta)$ Registration Errors}
    \label{tab:2}
\end{table}

\begin{figure}
    \centering
        \begin{minipage}[t]{0.45\textwidth}
            \includegraphics[width=\linewidth]{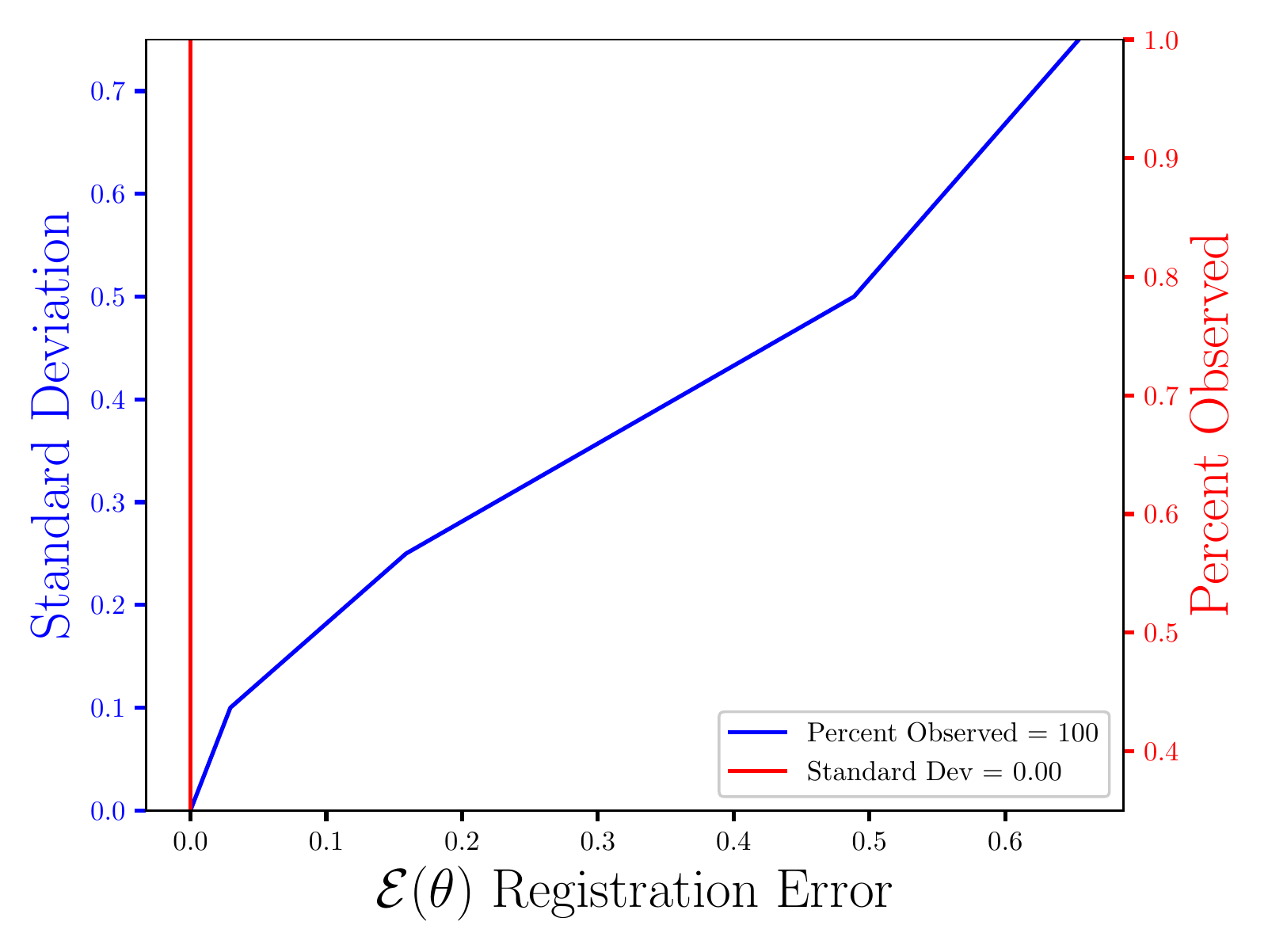}
            \caption{Blue: Full data, Red: Noiseless data (MALA)}\label{fig:mala1}
        \end{minipage}
        \hspace{0.5em}
        \begin{minipage}[t]{0.45\textwidth}
            \includegraphics[width=\linewidth]{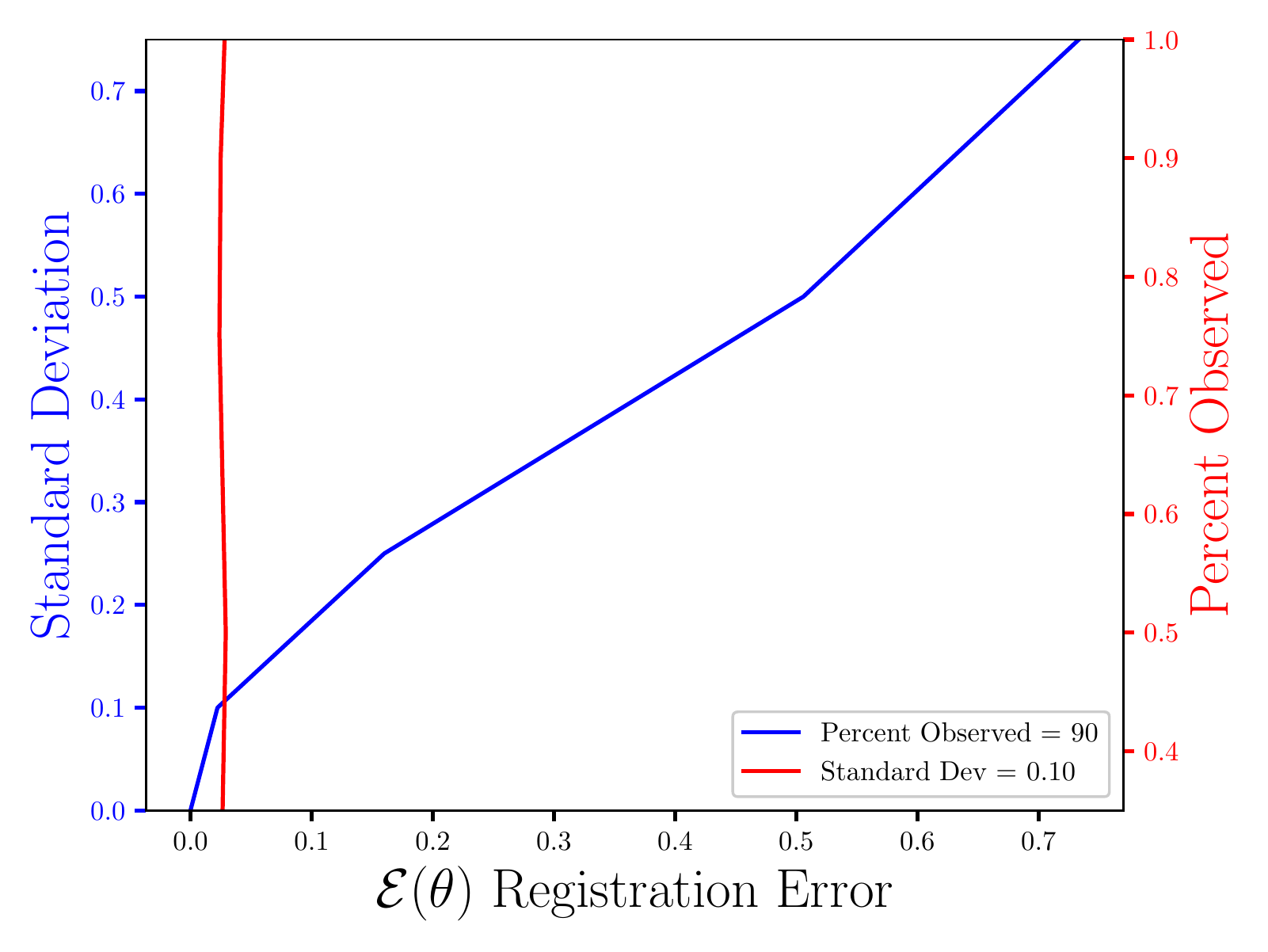}
            \caption{Blue: 90\% Observed, Red: $\gamma = 0.1$ (MALA)}\label{fig:mala2}
        \end{minipage}
\end{figure}

Furthermore, the results shown in Figs.~(\ref{fig:error1} - \ref{fig:error4}) are independent
of the algorithm, as the plots in Figs.~(\ref{fig:mala1} - \ref{fig:mala2}) show. 
For the latter, we ran a similar experiment with 125 i.i.d.~observation sets, but
to compute the registration, we used the Metropolis Adjusted Langevin Algorithm (MALA)
\cite{roberts1998optimal},
as opposed to HMC in Figs.~(\ref{fig:error1} - \ref{fig:error4}). Both
algorithms solve the same problem  and use information from the gradient of the 
log density. In the plots shown in Figs.~(\ref{fig:error1} - \ref{fig:error4}), 
we see the same constant error with
respect to the percent observed and the error increasing with the noise,
for a fixed percent observed. 
The MSE also appears to be proportional to $\gamma^2$, which is expected,
until some saturation threshold of $\gamma\geq 0.5$ or so.  
This can be understood as a threshold beyond
which the observed points will tend to get assigned to the wrong reference point.

\begin{figure}
    \centering
        \includegraphics[width=\linewidth]{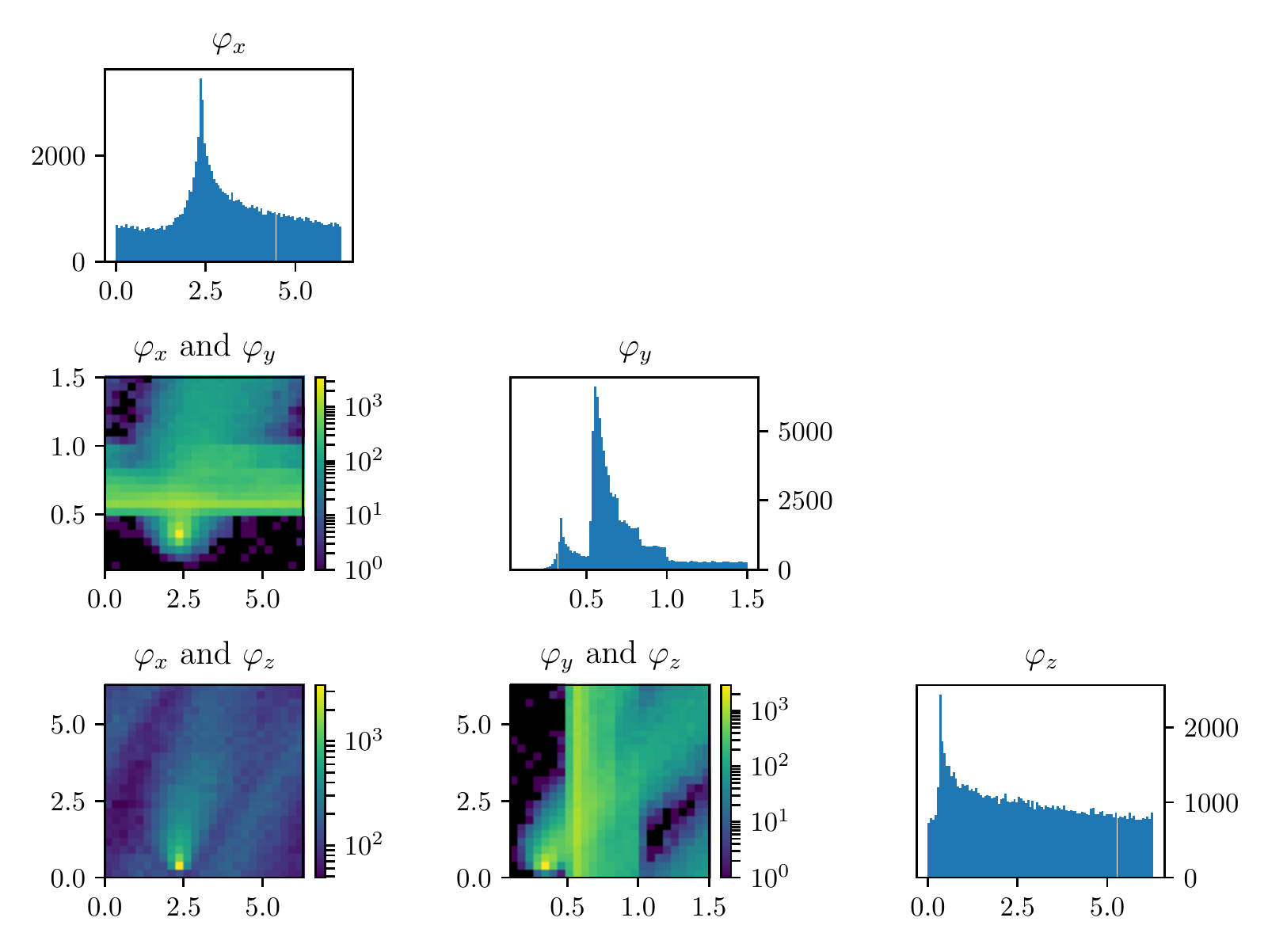}
        \caption{Histograms of $\varphi$ parameters, 100000 samples, $\gamma=0.25$, Observed = 35\%}\label{fig:angle_hist}
\end{figure}

\begin{figure}
    \centering
        \includegraphics[width=\linewidth]{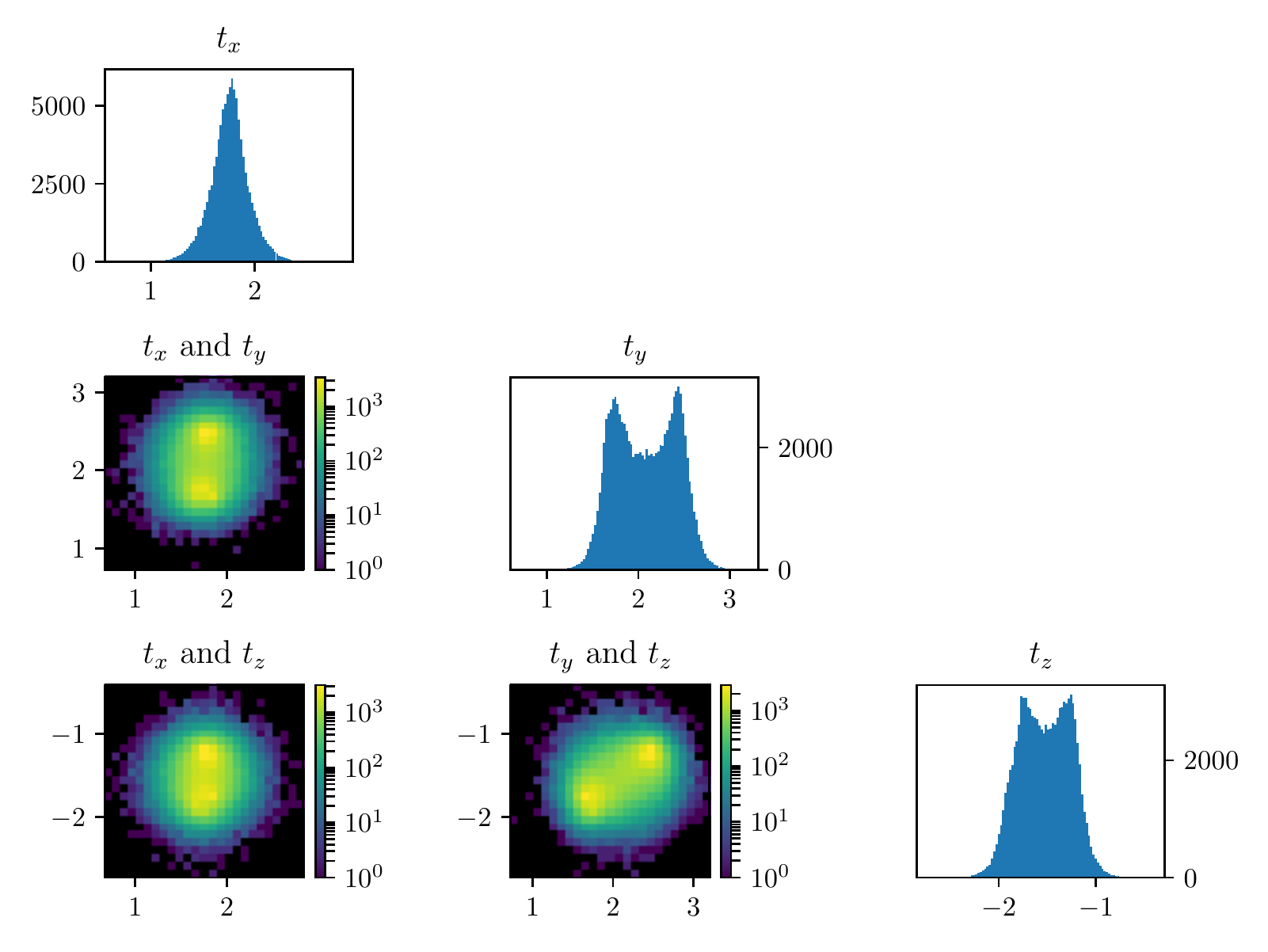}
        \caption{Histograms of $\theta$ parameters, 100000 samples, $\gamma=0.25$, Observed = 35\%}\label{fig:mcmc_hist1}
\end{figure}

\begin{figure}
    \centering
        \includegraphics[width=\linewidth]{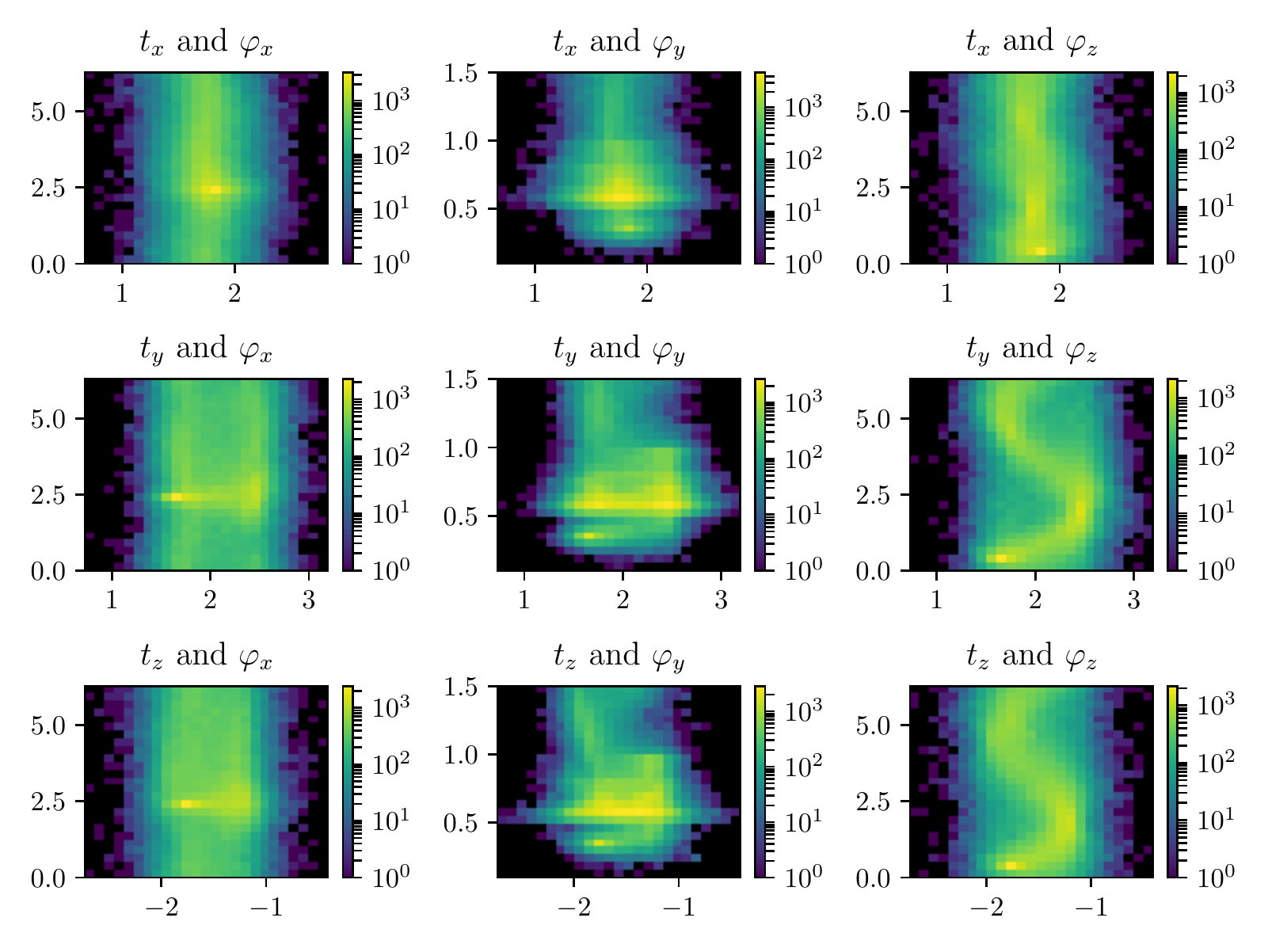}
        \caption{Histograms of $\theta$ parameters, 100000 samples, $\gamma=0.25$, Observed = 35\%}\label{fig:mcmc_hist2}
\end{figure}

To examine the contours of our posterior described by Eqn. (\ref{eqn:posterior}),
we drew $10^5$ samples from the density using the HMC methodology described previously. For this
simulation we set the noise to have standard deviation of 0.25
and the percent observed was 35\%, similar values to what we expect
from real APT datasets.
The rotation matrix $R$ is constructed via
Euler angles denoted: $\varphi_x, \varphi_y, \varphi_z$,
where $\varphi_x\in[0, 2\pi), \varphi_y\in[-\frac{\pi}{2}, \frac{\pi}{2}]$ and
$\varphi_z\in[0, 2\pi).$
These parameters are especially important to making
the correct atomic identification, which is crucial to the success of our method.

\begin{figure}
        \begin{minipage}[t]{0.45\textwidth}
            \includegraphics[width=\linewidth]{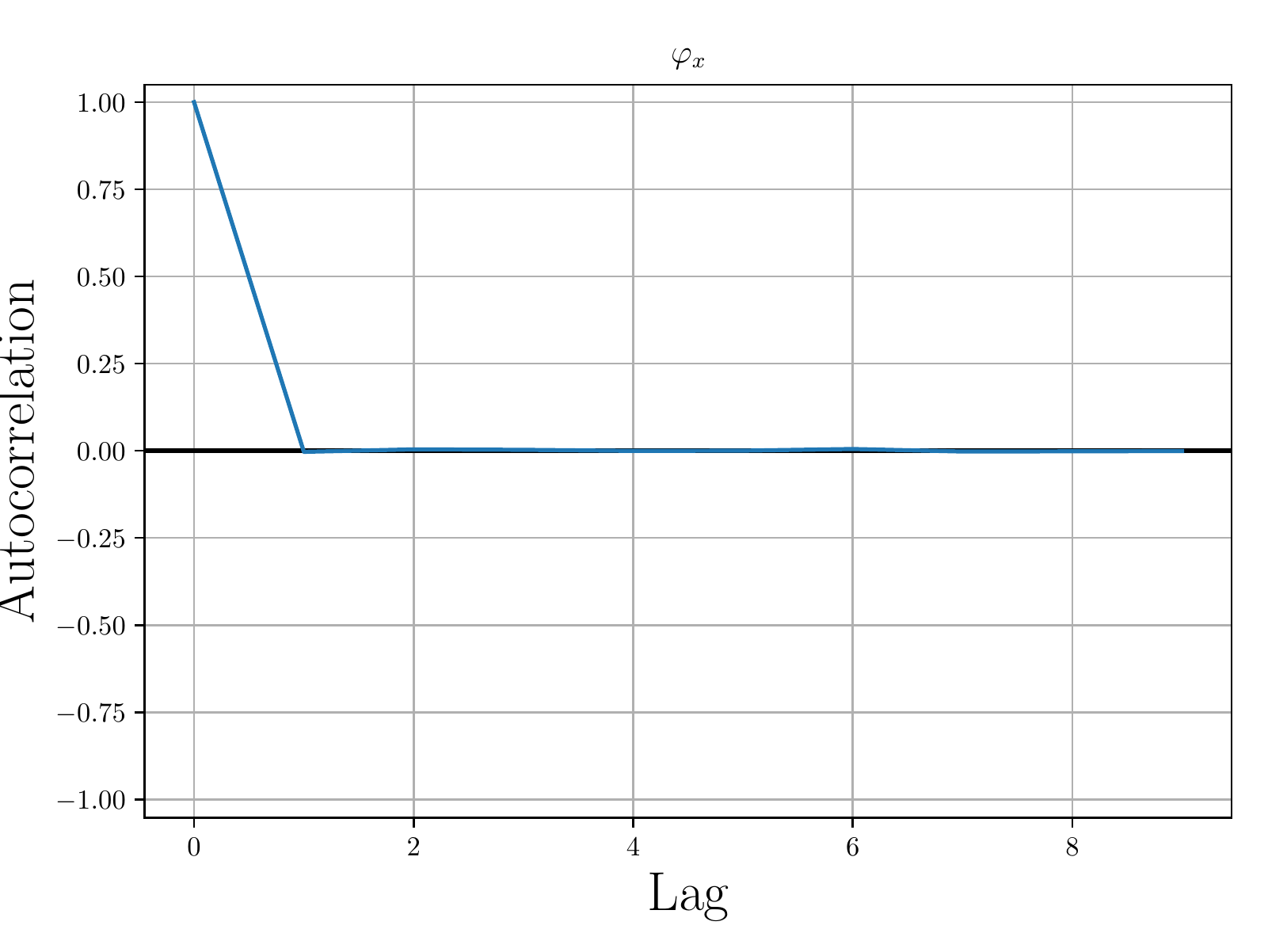}
            \captionof{figure}{Autocorrelation plot, $\varphi_x$}\label{fig:auto_corr1}
        \end{minipage}
        \hspace{0.5em}
        \begin{minipage}[t]{0.45\textwidth}
            \includegraphics[width=\linewidth]{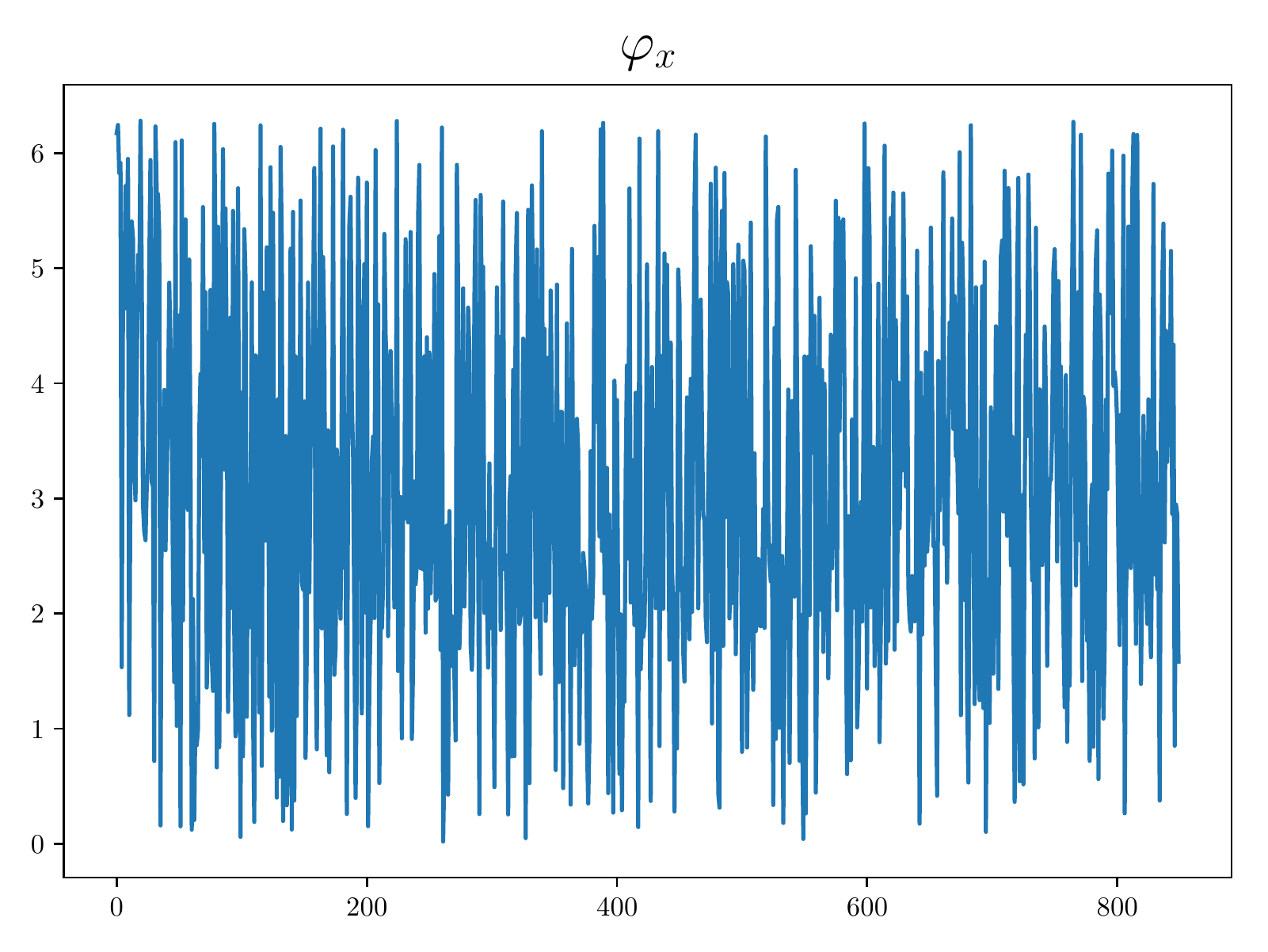}
            \captionof{figure}{Trace plot, $\varphi_x$}\label{fig:trace1}
        \end{minipage}
\end{figure}

\begin{figure}
        \begin{minipage}[t]{0.45\textwidth}
            \includegraphics[width=\linewidth]{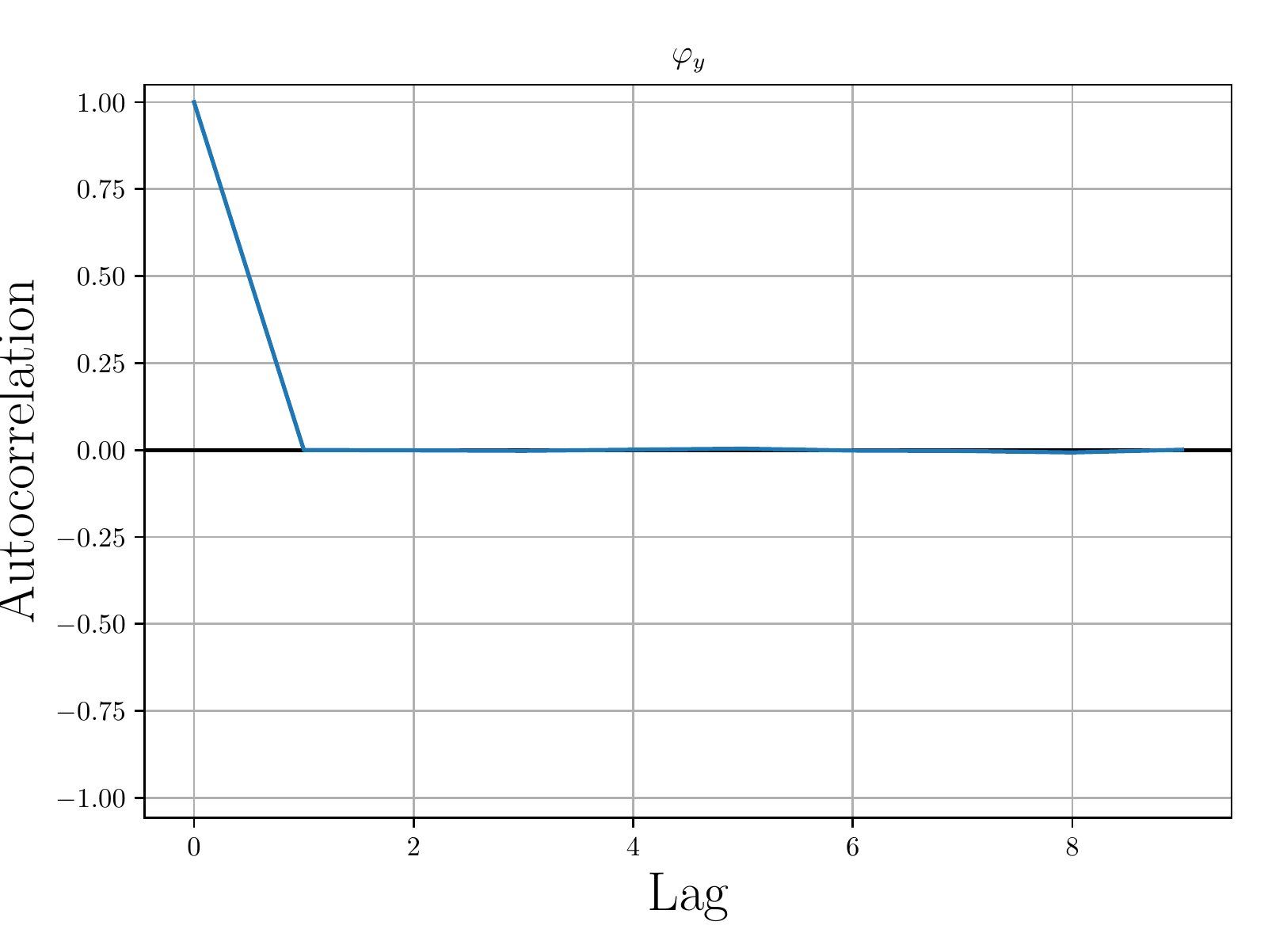}
            \caption{Autocorrelation plot, $\varphi_y$}\label{fig:auto_corr2}
        \end{minipage}
        \hspace{0.5em}
        \begin{minipage}[t]{0.45\textwidth}
            \includegraphics[width=\linewidth]{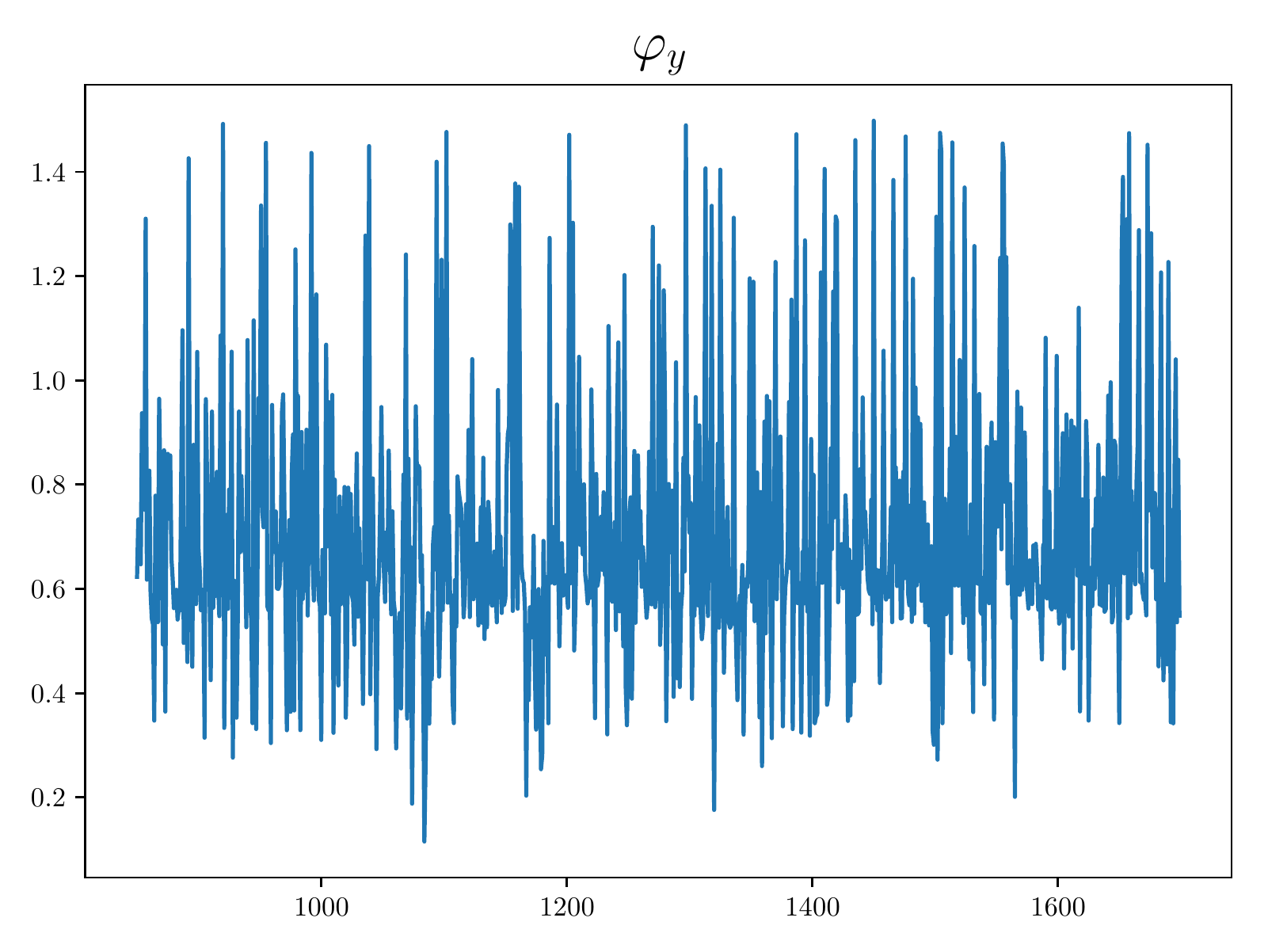}
            \caption{Trace plot, $\varphi_y$}\label{fig:trace2}
        \end{minipage}
\end{figure}

\begin{figure}
    \begin{center}
        \begin{minipage}[t]{0.45\linewidth}
            \includegraphics[width=\linewidth]{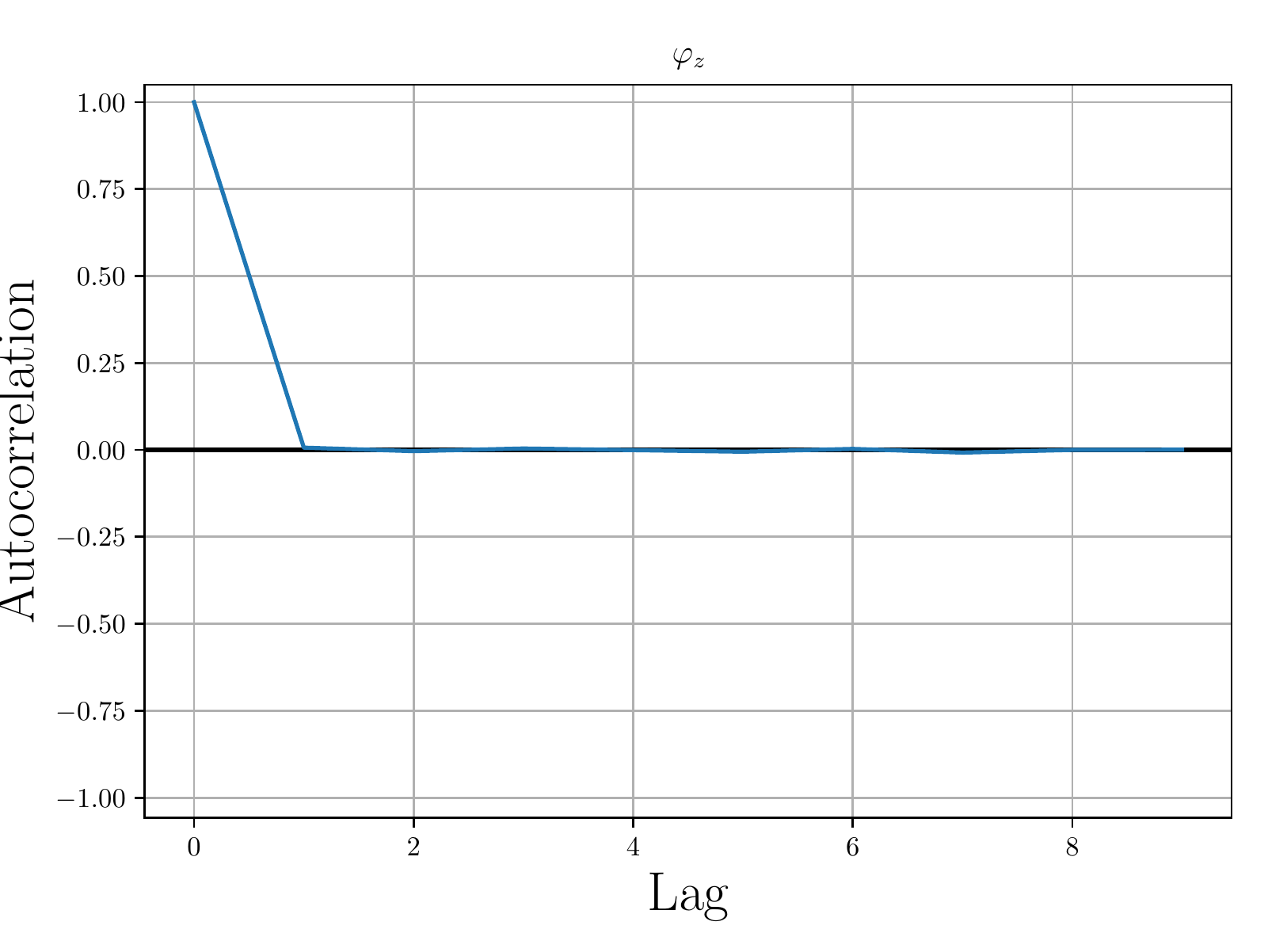}
            \caption{Autocorrelation plot, $\varphi_z$}\label{fig:auto_corr3}
        \end{minipage}
        \hspace{0.5em}%
        \begin{minipage}[t]{0.45\textwidth}
            \includegraphics[width=\linewidth]{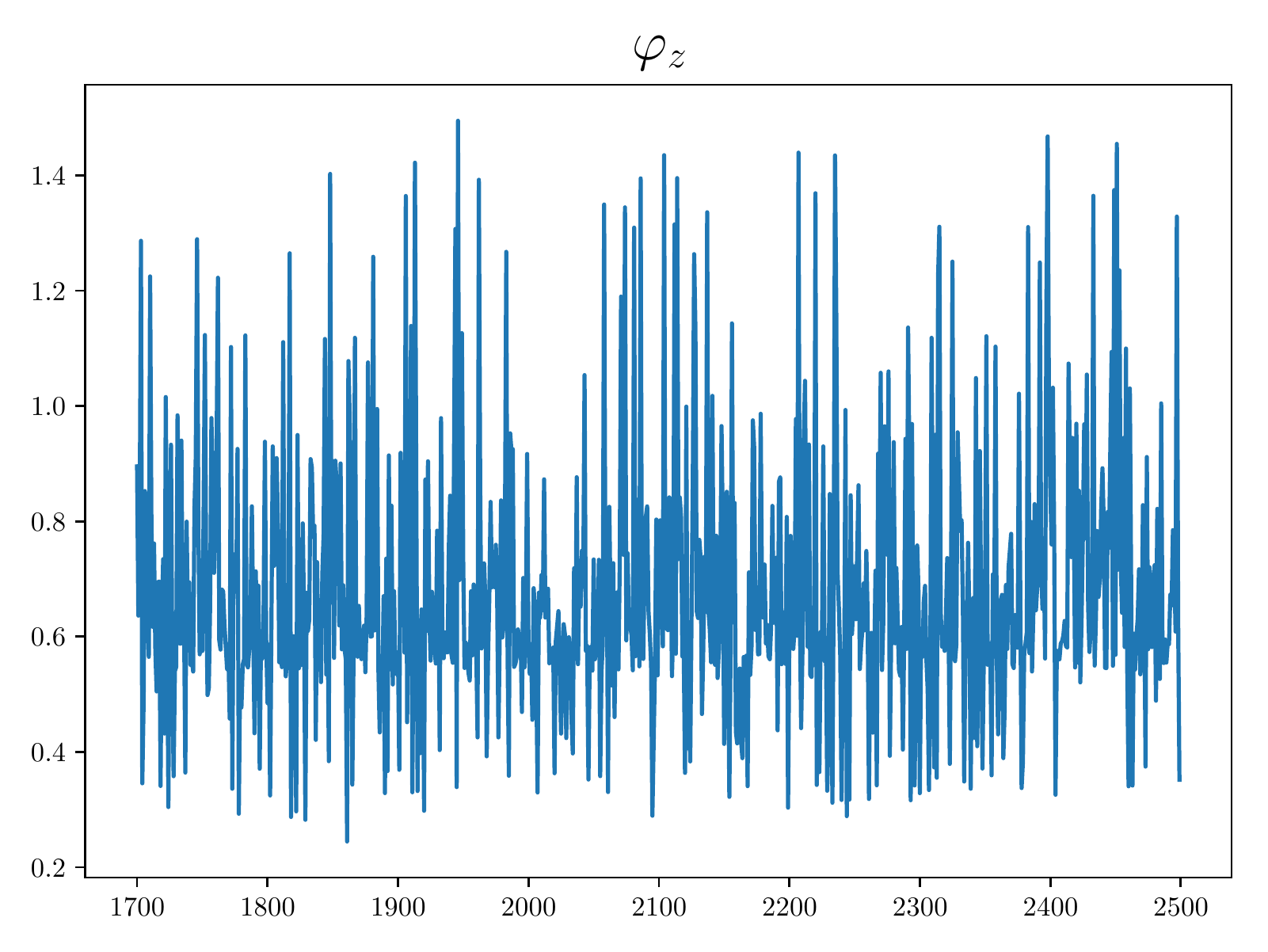}
            \caption{Trace plot, $\varphi_z$}\label{fig:trace3}
        \end{minipage}
    \end{center}
\end{figure}

In Figs.~(\ref{fig:angle_hist} - \ref{fig:mcmc_hist2}), 
we present marginal single variable histograms and all combinations
of marginal two-variable joint histograms for the individual components of $\theta$. 
We observe multiple modes in a number of the marginals.
In Figs. (\ref{fig:auto_corr1} - \ref{fig:trace3}) we present autocorrelation and trace plots
for the rotation parameters from the same instance of 
the HMC algorithm as presented in the histograms above in Figs.~(\ref{fig:angle_hist} - \ref{fig:mcmc_hist2}). 
We focus specifically on the rotation angles, to ensure efficient mixing of the Markov
chain as these have thus far been more difficult for the algorithm to optimize. 
We see the chain is mixing well with respect to these parameters
and appears not to become stuck in local basins of attraction.

Additionally, we consider the following. Define null sets $A_1, \dots, A_N$. 
For each $j=1,\dots,M$ and $l=1,\dots,L$, 
let $i^*(j,l) := {\rm argmin}_{i\in\{1,\dots,N\}} |R_{\varphi^l}^T(Y^l_j - t^l) - X_i|^2$,
and increment $A_{i^*(j,l)} = A_{i^*(j,l)} \cup Y^l_j$.
This provides a distribution of registered points for each index $i$, $A_i$, 
from which we 
estimate various statistics such as mean and variance. 
However, note that the cardinality 
varies between $|A_i| \in \{0,\dots,L\}$.
We are only be concerned with statistics around reference points $i$ 
such that $|A_i|>L/10$ or so, assuming that the other reference points 
correspond to outliers which were registered to by accident.  
{Around each of these $N'\leq N$ reference points 
$X_i$, we have a distribution of some $K\leq L$ registered points. 
We then computed the mean of these $K$ points, denoted by $\bar{X}_i$
and finally we compute the MSE $\frac1{N'}\sum_{i=1}^{N'} |X_i - \bar{X}_i|^2$.
The RMSE is 
reported in Table~\ref{tab:5}.
Here we note that a lower percentage observed $p$ is correlated with a larger error.
Coupling correct inferences about spatial alignment with
an ability to find distributions of atoms around each lattice point 
is a transformative tool for understanding High Entropy Alloys.

\begin{table}
    \caption{Errors for 125 Completed Registrations}
    \label{tab:5}
    \begin{tabular}{p{1.5cm}p{1.5cm}p{3.2cm}}
        \hline\noalign{\smallskip}
        Standard Deviation & Percent Observed &  Error  \\
        \noalign{\smallskip}\hline\noalign{\smallskip}
        0.25 &  75\% &  0.04909611134835241\\
        0.5 & 75\% & 0.07934531875006196 \\
        0.25 & 45\% &  0.07460005923988245\\
        0.5 & 45\% & 0.11978598998930728 \\
        \noalign{\smallskip}\hline\noalign{\smallskip}
    \end{tabular}
\end{table}

\section{Conclusion}\label{S:5}

We have presented a statistical model and
methodology for point set registration. We are able to recover a good estimate 
of the correspondence and spatial alignment between point sets in $\R^2$ and $\R^3$
despite missing data and added noise. As a continuation of this work,
we will extend the Bayesian framework presented in section
to incorporate the case of an unknown reference. In such a setting, we will seek not only
the correct spatial alignment and correspondence, but the reference point set, 
or crystal structure. The efficiency of our algorithm could be improved 
through a tempering scheme, allowing for easier transitions between modes, or an adaptive HMC 
scheme, where the chain learns about the sample space in order to make more efficient moves. 

Being able to recover the alignment and correspondences with an unknown reference 
will give Materials Science researchers an unprecedented tool
in making accurate predictions about High Entropy Alloys and allow them to
develop the necessary tools for classical interaction potentials.
Researchers working in the field will be able to determine the 
atomic level structure and chemical ordering of High Entropy Alloys. From such 
information, the Material Scientists will have the necessary tools to develop interaction
potentials, which is crucial for molecular dynamics simulations 
and designing these complex materials.

\section*{Acknowledgments}
A.S. would like to thank ORISE as well as Oak Ridge National Laboratory (ORNL)
Directed Research and Development funding. In addition, he thanks the 
CAM group at ORNL for their hospitality.  K.J.H.L. gratefully acknowledges 
the support of Oak Ridge National Laboratory Directed Research and Development funding.

\bibliographystyle{siam}
\bibliography{refs}

\end{document}